\documentclass[preprint,aps,pra,superscriptaddress]{revtex4-1}

\usepackage{subcaption}
\usepackage[normalem]{ulem}

\usepackage{lipsum}
\usepackage{amsfonts}
 \usepackage{amsmath}
\usepackage{amssymb}
 \usepackage{color}
\usepackage{graphicx}
\usepackage{epstopdf}
\usepackage{algorithmic}
\ifpdf
  \DeclareGraphicsExtensions{.eps,.pdf,.png,.jpg}
\else
  \DeclareGraphicsExtensions{.eps}
\fi

{\vskip 2pt \begin{compactitem}[#1]\setlength{\itemsep}{2pt}}
{\end{compactitem}\vskip 2pt}

\newcommand{\be}{\begin{equation}}
\newcommand{\ee}{\end{equation}} 

\newcommand{\OL}{\overline}

\newcommand{\br}{{\bf r}}
\newcommand{\bu}{{\bf u}}

\newcommand{\bg}{{\bf g}}

\newcommand{\bx}{{\bf x}}

\newcommand{\bomega}{{\mbox{\boldmath $\omega$}}}

\begin{document}

\title{Evolution of Rayleigh-Taylor turbulence under vorticity and strain-rate control}

\author{Dongxiao Zhao}
\author{Gaojin Li}
\email[]{\texttt{gaojinli@sjtu.edu.cn}}
\affiliation{State Key Laboratory of Ocean Engineering, School of Ocean and Civil Engineering, Shanghai Jiao Tong University, Shanghai 200240, China}

% \date{\today}

\begin{abstract}
We investigate the role of small-scale structures in turbulent Rayleigh-Taylor (RT) flows through the application of preferential flow control targeting high vorticity or high strain-rate regions \citep{Buzzicotti20PRL}. Through numerical simulations, we analyze the effects of flow control on RT statistics, mixing, and anisotropy behavior. Our results reveal that eliminating intense small-scale motion leads to the formation of more organized and coherent flow structures, with reduced mixing and enhanced anisotropy.  The alignment of vorticity and scalar gradient with the strain-rate eigen-frame is also altered by the flow control, reducing the downscale cascade of kinetic energy and the scalar variance. When the control threshold is set below the spatial mean of the vorticity or strain-rate field, turbulent motion in RT is significantly suppressed. Moreover, flow control eliminates regions of extreme vorticity and strain-rate, leading to overlapped high vorticity and high strain-rate regions with reduced turbulence intensity and more coherent structures. 
These findings provide a deeper understanding of the fundamental mechanisms played by small-scale structures in RT flows and their modulation through flow control. This work has broader implications for realistic scenarios, such as RT flows under magnetic fields or rotation, where suppression of small-scale motions plays a critical role. 

\end{abstract}

\maketitle

\section{Introduction}
The Rayleigh-Taylor (RT) instability is a fundamental hydrodynamic phenomenon that occurs when a denser fluid is accelerated against a lighter fluid, resulting in the formation of coherent structures such as bubbles and spikes, enhanced mixing, and eventually turbulent flows \citep{Zhou17-1,Zhou17-2,Boffetta17,Livescu20,Zhouetal2021Review}. This instability plays a pivotal role in a wide range of scientific and engineering phenomena, including supernova explosion \citep{Supernova00,Wang01,Cabot06}, inertial confinement fusion \citep{BettiHurricane16Nature,zhang2018nonlinear,campbell2021direct}, and combustion \citep{Ashurst96,Keenan2014IJHE,Sykes2021ProCI}. At its core, RT instability arises due to baroclinic torque generated by the misalignment of pressure and density gradients, thereby generating vorticity. Vorticity is not merely a byproduct of RT instability but a critical agent in shaping the flow. It governs the formation and evolution of coherent structures, influences energy transfer across scales, and determines the degree of anisotropy in RT turbulence. The amplification and redistribution of vorticity are central to the development of bubbles and spikes, directly impacting mixing efficiency and turbulence statistics.  While the statistical and temporal evolution of vorticity in RT turbulence has been extensively studied, it is often treated as a diagnostic tool to infer the flow state, mixing width growth, and anisotropy \citep{Zhou_2024,Qi2024PoF}. However, its direct dynamical role in the evolution of RT instability remains under-explored.

In a recent study, Buzzicotti, Biferale, and Toschi \citep{Buzzicotti20PRL} proposed a novel active control mechanism by modifying the incompressible Navier-Stokes equations to include a drag term that preferentially acts on regions of high vorticity. 
The drag term, which could potentially be implemented using Lagrangian particles such as self-controlled small magnetic objects, leverages the spinning-induced drag in high vorticity regions within a magneto-rheological fluid \citep{Stanway04, FalconPRF17, Buzzicotti20PRL}. Mathematically, it is expressed as:
\begin{align} \label{eq:control_incompr}
    f_c = c(\mathbf{x}, t) \mathbf{u}(\mathbf{x}, t),
\end{align}
where the spatially varying coefficient $c(\mathbf{x}, t)$ is defined as:
\begin{align} \label{eq:control_coeff_incompr}
    c(\mathbf{x}, t) = A_c \left\{1 + \tanh\left(|\boldsymbol{\omega}(\mathbf{x}, t)| - \omega_p\right)\right\}/2.
\end{align}
Here, the vorticity $\boldsymbol{\omega} = \nabla \times \mathbf{u}$, and $|\boldsymbol{\omega}(\mathbf{x}, t)|$ represents its magnitude. The control threshold $\omega_p$ is defined as $\omega_p = p \, \max \{ |\boldsymbol{\omega}| \}$, where $0 < p \leq 1$ is a constant parameter that determines the relative intensity of the control, and $A_c$ sets the overall magnitude of the drag force. Consequently, the strength of the control is governed by the two key parameters $p$ and $A_c$, which allow for precise tuning of the control mechanism.
Buzzicotti, Biferale, and Toschi \citep{Buzzicotti20PRL} demonstrated that this active control scheme effectively reduces intermittency, suppresses non-Gaussian statistics, and increases drag coefficients in isotropic turbulence. Remarkably, these effects are achieved even though the forcing is active only within a small volume fraction of the flow domain. This highlights the efficiency of the proposed control mechanism.

While the vorticity-control scheme offers a cost-effective approach to turbulence reduction by leveraging specific devices as discussed in Ref.~\cite{Buzzicotti20PRL}, it can also serve as a useful diagnostic tool for investigating the role of intense small scale structures such as vorticity in the evolution of turbulence. In this study, we adopt the latter perspective, utilizing the vorticity-control scheme to gain deeper insights into the dynamics of RT turbulence and the mechanisms driving its evolution.

Extending the above control scheme to RT turbulence offers a promising approach for exploring the role of vorticity in driving mixing, energy transfer, and anisotropy. By applying the drag term preferentially to high-vorticity regions, we could gain valuable insights into how vorticity influences the evolution of RT flows and the development of coherent structures. Furthermore, the proposed forcing mechanism could be adapted to target the magnitude of the strain-rate tensor, enabling a direct comparison between vorticity-controlled and strain-rate-controlled RT evolutions. Such studies would reveal the underlying mechanisms governing the interplay between vorticity and strain-rate dynamics in turbulent RT flows.

The paper is centered around two primary objectives. First, we aim to establish a fundamental understanding of the role that vorticity plays in the evolution of RT turbulence. This involves a comprehensive examination of both the qualitative and quantitative changes in turbulent statistics throughout the course of controlled RT evolution. Additionally, we will conduct a detailed comparison of the distinct effects that arise from vorticity-controlled and strain-rate-controlled RT dynamics, unveiling the unique contributions of each factor to the overall flow behavior.
Second, our focus extends to elucidating the dynamics underlying the proposed control scheme. Specifically, we seek to understand how this scheme suppresses the growth of the mixing width and mitigates the mixing induced by RT processes. External factors, such as magnetic fields or rotation, in RT flows usually play a significant role in modulating the flow dynamics and suppressing small-scale motions, which may share similarities with the current investigation. Thus the paper could also provides valuable insights into the real-world scenarios such as in inertial confinement fusion and astrophysical plasmas.

To achieve the stated objectives, this paper addresses the following key questions:\\
(1) How does the suppression of vorticity impact the evolution of the RT mixing layer and its anisotropy? \\
(2) Under the current control scheme, at what point is the RT turbulence fully suppressed? \\
(3) How do extreme events characterized by high vorticity and high strain influence the development of coherent structures in RT turbulence? 
Addressing these questions will yield deeper physical insights into the small-scale dynamics of RT turbulence and provide potential applications in a wide range of scientific and engineering contexts.

This paper is organized as follows. Section \ref{sec:equations} presents the governing equations of compressible RT instability, along with the numerical methods employed and the control schemes designed for RT suppression. 
Section \ref{sec:results} presents the results of the study. It emphasizes the key differences in RT growth, mixing behavior, and anisotropy between cases of baseline RT and those with flow control applied. Additionally, the small-scale alignment among vorticity, strain rate, and scalar gradient is thoroughly examined, shedding light on the intricate flow dynamics. Moreover, this section delves into the underlying dynamics that govern the effectiveness of the flow control mechanisms, offering a deeper understanding of the physical processes at play.
Finally, Section \ref{sec:conclusions} summarizes the main findings of the paper and discusses the practical implications of the current investigation.

\section{Governing equations and numerical methods} \label{sec:equations}
The governing equations are formulated based on the conservations of mass, mass fraction, momentum, and energy. To incorporate the effects of flow control, a forcing term is introduced into the momentum equation, specifically designed to target regions characterized by intense vorticity or strain. The equations in dimensionless form are \cite{Zhao22JFM,ZhaoLi25}
\begin{eqnarray} \label{eq:compressible}
    &\frac{\partial \rho}{\partial t} + \nabla \cdot (\rho \bu) = 0 \label{eq:mass}\\
    &\frac{\partial \rho Y}{\partial t} + \nabla\cdot (\rho \bu Y) = \frac{1}{\mathrm{Re_S}\mathrm{Sc}}\nabla\cdot(\rho \nabla Y) \label{eq:mass_fraction}\\    
    &\frac{\partial \rho \bu}{\partial t} + \nabla\cdot(\rho \bu \bu) = -\nabla P+\frac{1}{\mathrm{Re_S}}\nabla\cdot 2\left(\mathbf{S} - \frac{1}{3}(\mathbf{S} : \mathbf{I})\right) - \frac{1}{\mathrm{Fr}}\rho \delta_{iz} - \boldsymbol{f}_c\label{eq:momentum}\\
    &\frac{\partial \rho E}{\partial t} + \nabla\cdot ((\rho E + P)\bu) = \frac{1}{\mathrm{Re_S}}\nabla\cdot (2\left(\mathbf{S} - \frac{1}{3}(\mathbf{S} : \mathbf{I})\right)\cdot \bu) - \frac{1}{\mathrm{Fr}} \rho u_z + \frac{1}{\mathrm{Re_S}\mathrm{Pr}}\nabla\cdot (\nabla T) \label{eq:energy}
\end{eqnarray}
% \begin{eqnarray} \label{eq:compressible}
%     &\frac{\partial \rho}{\partial t} + \nabla \cdot (\rho \bu) = 0 \label{eq:mass}\\
%     &\frac{\partial \rho Y}{\partial t} + \nabla\cdot (\rho \bu Y) = \nabla\cdot(\rho D \nabla Y) \label{eq:mass_fraction}\\    
%     &\frac{\partial \rho \bu}{\partial t} + \nabla\cdot(\rho \bu \bu) = -\nabla P+\nabla\cdot \boldsymbol{\tau} +\rho \boldsymbol{g} - \boldsymbol{f}_c\label{eq:momentum}\\
%     &\frac{\partial \rho E}{\partial t} + \nabla\cdot ((\rho E + P)\bu) = \nabla\cdot (\boldsymbol{\tau}\cdot \bu) + \rho \bu\cdot \boldsymbol{g} + \nabla\cdot (\kappa \nabla T) \label{eq:energy}
% \end{eqnarray}
Here, $\rho$, $\mathbf{u}$ and $P$ represent the non-dimensional density, velocity and pressure, respectively. The reference length $L_\mathrm{ref}$, reference density $\rho_\mathrm{ref}$, and reference pressure $P_\mathrm{ref}$ are used for non-dimensionalization. $Y$ denotes the mass fraction of the heavy fluid, and $E = \frac{1}{2}u^2 + c_v T$ is the total energy density, where $c_v$ is the specific heat at constant volume and $T$ is the temperature.  
The strain rate tensor is defined as $\mathbf{S} = (\nabla \mathbf{u} + \nabla \mathbf{u}^T)/2$ and $\boldsymbol{I}$ is the identity tensor. 
The non-dimensional quantities are given by the static Reynolds number, the Froude, Prandtl, and Schmidt numbers 
\begin{align}
    \mathrm{Re_S} = \frac{\rho_\mathrm{ref}U_\mathrm{ref}L_\mathrm{ref}}{\mu}, \; \mathrm{Fr}=\frac{U_\mathrm{ref}^2}{gL_\mathrm{ref}}, \; \mathrm{Pr} = c_p \frac{\mu}{\kappa}, \; \mathrm{Sc} = \frac{\mu}{\rho_\mathrm{ref} D}
\end{align}
The parameters $\mu$, $D$, and $\kappa$ represent the dynamic viscosity, mass diffusivity, and thermal conductivity, respectively, and $c_p$ is the specific heat at constant pressure.
The set of equations (\ref{eq:mass})-(\ref{eq:energy}) is closed using the ideal gas equation of state, $P = \rho \widetilde{R} T \left(\frac{Y}{W_h} + \frac{1 - Y}{W_l}\right)$, where $\widetilde{R}$ is the universal gas constant, and $W_h$ and $W_l$ are the molecular weights of the heavy and light fluids, respectively. Note that the above compressible formulation in equations (\ref{eq:mass})-(\ref{eq:energy}) reduces to the two-species variable density incompressible formulation in the limit of zero Mach number \citep{Sandoval95}.

Following reference \citep{Buzzicotti20PRL}, the friction or drag term of the active control in Eq.~(\ref{eq:momentum}) is defined as 
\begin{align} \label{eq:control}
   \boldsymbol{f}_c = c(\mathbf{x}, t)\rho\mathbf{u}(\mathbf{x}, t) 
\end{align}
where the coefficient $c(\mathbf{x}, t)$ takes different forms depending on the control mechanism. For vorticity-based control, it is given by  
$c(\mathbf{x}, t) = A_c \{1 + \tanh\left(|\boldsymbol{\omega}(\mathbf{x}, t)| - \omega_p\right)\}/2$,  
while for strain-rate-based control, it is defined as  
$c(\mathbf{x}, t) = A_c \{1 + \tanh\left(|\mathbf{S}(\mathbf{x}, t)| - S_p\right)\}/2$.
The control thresholds $\omega_p$ and $S_p$ are expressed as $\omega_p = p \, \mathrm{max} \{|\boldsymbol{\omega}|\}$ and $S_p = p \, \mathrm{max} \{|\mathbf{S}|\}$, respectively. For the parameters $p$ and $A_c$, we vary only the control threshold parameter $p$, while keeping the control magnitude $A_c$ fixed according to $A_c \equiv \sqrt{\mathcal{A} g / L_c}$. Here, $L_c$ corresponds to the characteristic length scale associated with the peak of the initial velocity perturbation spectrum, and $\mathcal{A} = (\rho_h - \rho_l) / (\rho_h + \rho_l)$ is the Atwood number, based on the initial uniform densities of the heavy and light fluids, $\rho_h$ and $\rho_l$.  The primary objective of introducing this flow control, which for most cases is active only within a small portion of the entire domain, is to serve as a diagnostic tool for investigating the influence of small-scale vorticity or strain on the overall evolution of the RT instability.

In the numerical simulations, the computational domain is a three-dimensional rectangular box with a non-dimensional size of $L_x \times L_y \times L_z = 0.8 \times 0.8 \times 1.6$. The domain is filled with heavy and light fluids in the top and bottom halves, with densities of $\rho_h = 1$ and $\rho_l = 0.739$, respectively. This configuration results in a Atwood number of $\mathcal{A} = (\rho_h - \rho_l)/(\rho_h + \rho_l) = 0.15$.   Periodic boundary conditions are applied in the two horizontal directions, while no-slip boundary conditions are imposed at the top and bottom walls.  The governing equations, (\ref{eq:mass})-(\ref{eq:energy}), are solved numerically using a pseudo-spectral method in the horizontal directions and a sixth-order compact finite difference scheme in the vertical direction. Time integration is performed using a fourth-order Runge-Kutta method. Further details of the numerical approach can be found in Ref.~\cite{Zhao22JFM}.

\section{Results and Discussion} \label{sec:results}

The simulation parameters used in this study are summarized in Table~\ref{tab:parameters}. We examined vorticity-controlled cases relative to the baseline (`Base') case, testing three different control threshold parameters with $p = 0.5$, $0.35$, and $0.2$.  We also tested cases with higher control parameter ($p=0.7$ and  $0.85$), but the results were statistically similar to the baseline case and thus are not included. Additionally, a strain-rate-controlled case with a parameter value of $p = 0.2$ was incorporated, along with a high-resolution baseline case for comparison. All simulations are verified to meet the criterion for resolving the smallest length scales \citep{YeungPope1989JFM}.

All simulations in Table~\ref{tab:parameters} maintain a low Mach number, confirming that the RT flows studied here are nearly incompressible. The Atwood number is fixed at $\mathcal{A} = 0.15$ across all cases. The simulations begin from identical initial conditions, where the vertical velocity field is perturbed with uniform fluctuations within the spectral range $k \in [16, 32]$ in wavenumber space. 
The vorticity- or strain-rate-based control is applied throughout the simulation, with thresholds $\omega_p \equiv p \, \mathrm{max} \{|\boldsymbol{\omega}|\}$ and $S_p \equiv p \, \mathrm{max} \{|\mathbf{S}|\}$ updated dynamically at each time step.

\begin{table} \centering
\setlength{\tabcolsep}{4pt}
    \caption{The parameters for the 3D RT simulations in this paper. In non-dimensional form, the domain size is $L_x, L_y, L_z = 0.8, 0.8, 1.6$, and the Atwood number is $\mathcal{A} = 0.15$. $\mathrm{Re}_S\equiv \rho_\mathrm{ref} U_\mathrm{ref}L_\mathrm{ref}/\mu$ is the static Reynolds number, while the outer-scale Reynolds number is defined as $Re = \langle \rho \rangle h \dot{h}/\mu$, where $\langle \cdot \rangle$ denotes the spatial mean, $h$ is the mixing width determined by the 5\%--95\% threshold of the mass fraction field, and $\mu$ is the dynamic viscosity. For all cases, the Prandtl number $\mathrm{Pr}$ and the Schmidt number $\mathrm{Sc}$ are set to unity, the Reynolds number and the maximum Mach number $\mathcal{M}$ are calculated at time steps corresponding to a mixing width of $h = 1.0$, and the control threshold are all set to $A_c=\sqrt{\mathcal{A}g/L_c}=1.97$.}
\begin{tabular}{|c|c|c|c|c|c|c|c|}
    \hline label & grid size &  $1/\mathrm{Re}_S$ & control & $p$ & $\mathrm{Re}$ & $\mathcal{M}$  \\ \hline
    Base & $256\times 256\times 512$& $2.5\times 10^{-5}$ & -- & -- & 11667 & $5.1\times 10^{-4}$  \\ \hline
    Refine & $512\times 512\times 1024$& $1.0\times 10^{-5}$ & -- & -- & 17662 & $1.2\times 10^{-3}$  \\ \hline
    LowRe & $256\times 256\times 512$& $3.75\times 10^{-5}$ & -- & -- & 6369 & $4.3\times 10^{-4}$  \\ \hline
    W05 &$256\times 256\times 512$& $2.5\times 10^{-5}$ & $|\bomega|$ & 0.5 & 10598 &  $3.8\times 10^{-4}$  \\ \hline
    W035 &$256\times 256\times 512$&  $2.5\times 10^{-5}$ & $|\bomega|$ & 0.35 & 9415 & $4.2\times 10^{-4}$  \\ \hline
    W02 &$256\times 256\times 512$&  $2.5\times 10^{-5}$ & $|\bomega|$ & 0.2 & 8491 &  $4.3\times 10^{-4}$  \\ \hline    
    % W02Fix &$256\times 256\times 512$&  $2.5\times 10^{-5}$ & $|\bomega|$ & 0.2 & 7808 &  0.15  \\ \hline
    % WnegInf &$256\times 256\times 512$&  $2.5\times 10^{-5}$ & $|\bomega|$ & $-\infty$ & 7808 &  0.15  \\ \hline
    S02 &$256\times 256\times 512$&  $2.5\times 10^{-5}$ & $|\boldsymbol{S}|$ & 0.2 & 6968 &  $3.6\times 10^{-4}$  \\ \hline
\end{tabular}
\label{tab:parameters}
\end{table}

\begin{figure}%[hbt!]
\centering
\begin{minipage}[b]{1.0\textwidth}
\centering
\begin{subfigure}{0.98\textwidth}  
    \centering
    \includegraphics[width=0.98\textwidth]{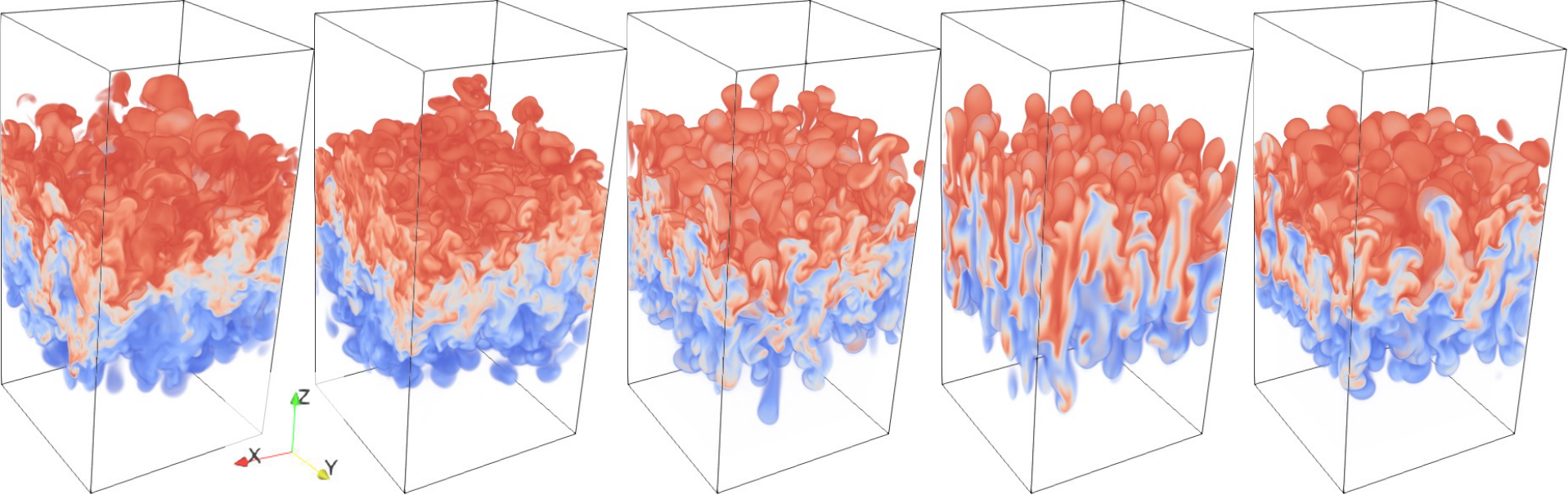}   
    %\caption{}
    \end{subfigure}     
    \begin{subfigure}{0.98\textwidth}  
    \centering
    \includegraphics[width=0.98\textwidth]{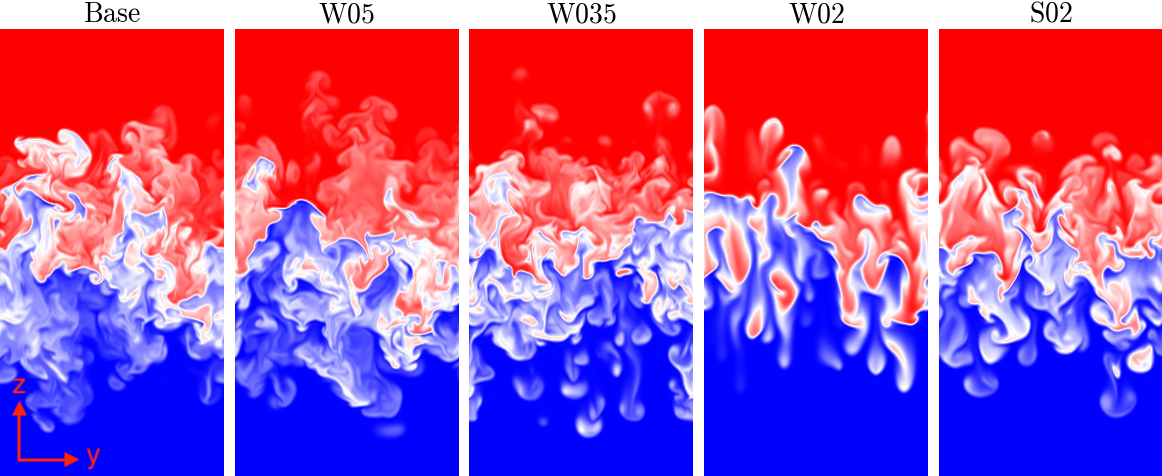}   
    %\caption{}
    \end{subfigure}        
\end{minipage}
    \caption{Visualizations of the 3D mass fraction fields (top) and at $y$-$z$ slices (bottom) are presented for both the baseline and controlled simulation cases, captured at the respective time instances when the mixing width $h = 1.0$. \label{fig:RT_viz}}
\end{figure}

Figure \ref{fig:RT_viz} presents 3D visualizations and 2D slices of the mass fraction field $Y$ for different simulation cases at the point where their mixing width reaches $h = 1.0$.  The mixing width quantifies the extent to which heavy and light fluids interpenetrate each other and is defined based on the 5\%-95\% heavy fluid mass fraction levels.
As the control threshold $p$ decreases from 0.5 (case W05) to 0.2 (case W02), the vorticity control is applied to a larger volume fraction of the domain, leading to stronger control effects. This intensified control results in a more organized flow pattern in case W02.  In contrast to the baseline case, where the heavy and light fluids exhibit chaotic mixing, the bubble and spike structures remain discernible in case W02, with minimal horizontal motion distorting these structures. The observed flow pattern suggests that the Kelvin-Helmholtz instability, typically generated at the interface between the heavy and light fluids and creating high vorticity, is effectively suppressed. This suppression reduces the interaction between coherent structures in the flow, leading to a lower transfer of mass and energy to smaller scales and, consequently, an overall lower turbulence level. 
A similar suppression effect is observed in the strain-rate-controlled case (S02). However, the degree of suppression is less pronounced compared to the vorticity-controlled case (W02) with the same control level.

In addition, the visualizations for the Refine and lowRe cases are presented in Fig.~\ref{Appfig:viz_refine_lowRe} within Appendix \ref{Appsec:viz}. When compared to the controlled cases, which selectively suppress the flow by targeting high vorticity or high strain-rate regions, the lowRe case employs a uniform increase in viscosity across the entire flow domain to decrease the turbulence level. 
Notably, despite the fact that the Reynolds number of the lowRe case is smaller than that of the W02 case, its flow pattern in Fig.~\ref{Appfig:viz_refine_lowRe} is more intricate and complex. The lowRe case exhibits substantial horizontal motion, which gives rise to complex flow structures. 
These observations  demonstrate that the preferential control approach is more efficient in suppressing RT turbulence than simply increasing the viscosity uniformly. 

In the following analysis, we will concentrate on comparing the baseline case with the controlled cases at different levels. The comparisons will encompass key aspects such as mixedness, anisotropy, and small-scale alignment statistics, providing a comprehensive understanding of the differences and similarities between these cases.

\subsection{Impact of vorticity control on RT temporal statistics}

We begin by examining the influence of the adopted flow control, which acts preferential on regions of high vorticity or strain-rate, on the statistical evolution of RT instability. Two key properties characterizing RT flows are the mixing width, $h(t)$, and the mixedness parameter, $\Theta$. 
In this study, the mixing width $h(t)$ is defined as the vertical distance between two locations where the horizontally averaged mass fraction of the heavy fluid reaches 5\% and 95\%, respectively \citep{Zhou17-1,Bian20}. The mixedness parameter $\Theta$ is expressed as  
\begin{align}
\Theta = \frac{\int_{-\infty}^\infty \langle Y(1 - Y) \rangle_h \, dz}{\int_{-\infty}^\infty \langle Y \rangle_h \langle 1 - Y \rangle_h \, dz},
\end{align}  
where $Y$ is the mass fraction of the heavy fluid. The angle brackets $\langle \cdot \rangle_h= \int (\cdot) \, dx \, dy$ denote averaging over the horizontal plane.

\begin{figure}%[hbt!]
\centering
\begin{minipage}[b]{1.0\textwidth}
\centering
    \begin{subfigure}{0.4\textwidth}  
    \centering
    \includegraphics[width=0.98\textwidth]{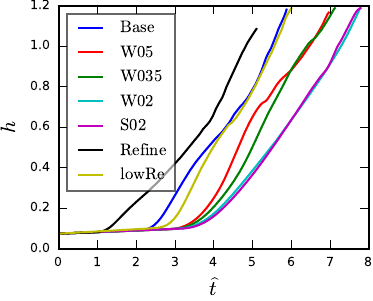}   
    \caption{}
    \end{subfigure} \\
    \begin{subfigure}{0.4\textwidth}  
    \centering
    \includegraphics[width=0.98\textwidth]{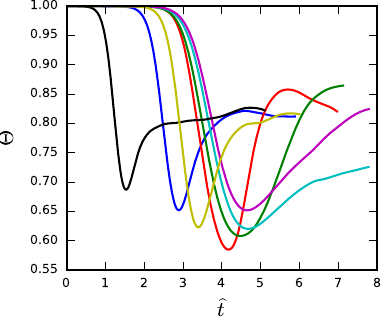}   
    \caption{}
    \end{subfigure} 
    \begin{subfigure}{0.4\textwidth}  
    \centering
    \includegraphics[width=0.98\textwidth]{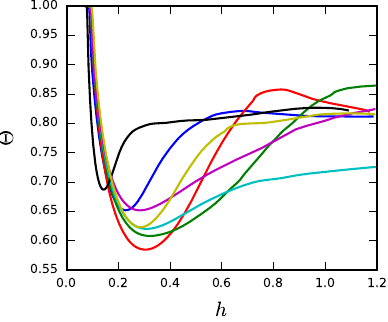}   
    \caption{}
    \end{subfigure} 
\end{minipage}
    \caption{(a) Time evolution of the mixing width for the cases illustrated in Fig.~\ref{fig:RT_viz}. The time variable is non-dimensionalized using the characteristic time scale, $\widehat{t} = t\sqrt{\mathcal{A}g/L_x}$, where $\mathcal{A}$ denotes the Atwood number. (b) Evolution of the mixedness parameter, $\Theta$, as a function of time. (c) Variation of $\Theta$ with respect to the mixing width $h$. \label{fig:width_theta}}
\end{figure}

Fig.~\ref{fig:width_theta} presents the growth of the mixing width and the evolution of the mixedness parameter across different cases. In Fig.~\ref{fig:width_theta}(a), the mixing width shows faster growth in the refined case with reduced viscosity and thermal diffusivity compared to the baseline case. In addition, growth is progressively delayed with reducing Reynolds number and with increasing levels of vorticity or strain control. 
The temporal evolution of the mixedness parameter, $\Theta$, shown in Fig.~\ref{fig:width_theta}(b), highlights phase differences across cases due to variations in mixing width growth. However, when $\Theta$ is plotted against $h$ in Fig.~\ref{fig:width_theta}(c), different cases exhibit more similarity during the initial stages of flow development, with differences becoming pronounced only in the later stages. This observation indicates that the mixing width $h$ serves as a more consistent and reliable parameter for comparing different RT evolutions compared to time. As a result, $h$ is chosen as a substitute for the time parameter in the subsequent analysis.
In Fig.~\ref{fig:width_theta}(c), it is further observed that all cases, except W02, asymptote to a constant $\Theta$ value slightly above 0.8, consistent with previously reported results \cite{Cook2004JFM, BanerjeeAndrews2009}. The W02 case, however, results in a lower asymptotic $\Theta$ value of approximately 0.723, reflecting a reduced degree of mixing. This finding aligns with the visualizations of the W02 case shown in Fig.~\ref{fig:RT_viz}.

\begin{figure}%[hbt!]
\centering
\begin{minipage}[b]{1.0\textwidth}
\centering    
    \begin{subfigure}{0.4\textwidth}  
    \centering
    \includegraphics[width=0.98\textwidth]{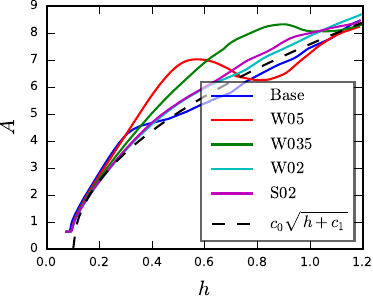}   
    \caption{}
    \end{subfigure} 
    \begin{subfigure}{0.4\textwidth}  
    \centering
    \includegraphics[width=0.98\textwidth]{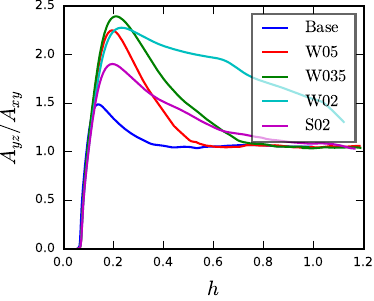}   
    \caption{}
    \end{subfigure} 
\end{minipage}
    \caption{(a) Total interface area as a function of the mixing width. An auxiliary line $c_0\sqrt{h+c_1}$ is included, where the parameters $c_1=0.1, c_0 = L_xL_y/\sqrt{c_1}=8$.  (b) Ratio of the interface area projected onto the $y$-$z$ (vertical) plane to that on the $x$-$y$ (horizontal) plane. \label{fig:area_ratio}}
\end{figure}

The degree of mixedness in turbulent flows is closely related to the wrinkle and distortion of the material interface. As the RT flow evolves, the interface between the heavy and light fluids becomes increasingly distorted and stretched, with its total area $A$ growing approximately linearly over time \citep{CabotCookNatPhys}, and hence $A(t)\propto \sqrt{h(t)}$. Figure~\ref{fig:area_ratio}(a) shows the evolution of the interface area as a function of the mixing width. The interface area is defined by the $Y = 0.5$ isosurface and calculated using the marching cubes algorithm \citep{MarchingCube}. Across all cases studied, the total interface area increases roughly as $A\propto \sqrt{h}$, and the magnitudes are similar when plotted against the mixing width. For the cases considered here, the area expands approximately 13-fold at $h = 1.2$ compared to the initial flat horizontal interface.

A more pronounced difference arises in the orientation of the interface, quantified by the ratio of the interface area projected onto the vertical ($y$-$z$) plane to that on the horizontal ($x$-$y$) plane, $A_{yz}/A_{xy}$, as shown in Fig.~\ref{fig:area_ratio}(b). In the baseline case, this ratio evolves from 0 to approximately 1.5 before settling into an isotropic state with $A_{yz}/A_{xy} \approx 1$. In contrast, for cases with increasing levels of vorticity or strain-rate control, $A_{yz}/A_{xy} > 1$ persists for a longer duration before eventually converging to 1. In W02, the ratio remains greater than 1 during the late stages of the flow; however, it exhibits a consistent decreasing trend, suggesting a gradual tendency toward 1.
Visualizations in Fig.~\ref{fig:RT_viz} reveal that in the W02 case, bubble and spike structures are sustained due to the suppression of the Kelvin-Helmholtz instability resulting from vorticity control. These coherent structures predominantly elongate in the vertical direction, resulting in a consistently high $A_{yz}/A_{xy}$ ratio. While this ratio does decrease with increasing $h$, indicating some development of horizontal structures in the density field, the vertically aligned structures continue to dominate throughout the simulated time span.

In addition to the density and mass fraction fields, the kinetic energy (KE) at large scales, characterized by the velocity field, and at small scales, characterized by strain rate and vorticity, are significantly affected by preferential control over regions of high vorticity or strain. Under such control, the budget equations for KE, squared vorticity (enstrophy), and squared strain rate are given as follows
\begin{align}
\frac{D}{Dt} \frac{|\bu|^2}{2} &= -\frac{1}{\rho}\bu\cdot \nabla P + \frac{1}{\rho} (\nabla\cdot \boldsymbol{\tau}) \cdot \bu +\bu\cdot\bg \underbrace{-\bu\cdot\boldsymbol{f}_c}_{\dot{\mathrm{KE}}_\mathrm{control}} \label{eq:KE}\\
     \frac{D}{Dt} \frac{|\boldsymbol{\omega}|^2}{2} &= \underbrace{- |\boldsymbol{\omega}|^2\nabla \cdot \boldsymbol{u}}_{\dot{\Omega}_\mathrm{dilate}} +  \underbrace{\boldsymbol{\omega}\cdot\nabla\boldsymbol{u}\cdot\boldsymbol{\omega}}_{\dot{\Omega}_\mathrm{stretch}} + \underbrace{\frac{\boldsymbol{\omega}}{\rho^2}\cdot (\nabla\rho\times\nabla P)}_{\dot{\Omega}_\mathrm{baro}} + \underbrace{\boldsymbol{\omega}\cdot \nabla\times \left(\frac{1}{\rho} \nabla\cdot \tau\right)}_{\dot{\Omega}_\mathrm{dissip}} \underbrace{-\boldsymbol{\omega}\cdot \nabla\times \frac{\boldsymbol{f}_c}{\rho}}_{\dot{\Omega}_\mathrm{control}} \label{eq:vort_sq}\\
     \frac{D}{Dt} \frac{|\boldsymbol{S}|^2}{2} &= \underbrace{-\boldsymbol{S}:(\nabla \boldsymbol{u} \nabla \boldsymbol{u})}_{\dot{\mathcal{S}}_\mathrm{amplify}} + \underbrace{\frac{1}{\rho^2} \nabla P \cdot \boldsymbol{S}\cdot \nabla \rho}_{\dot{\mathcal{S}}_\mathrm{baro}} \underbrace{- \frac{1}{\rho}\nabla\nabla P: \boldsymbol{S}}_{\dot{\mathcal{S}}_\mathrm{Hessian}} + \underbrace{\boldsymbol{S}:\nabla \left(\frac{1}{\rho} \nabla\cdot \boldsymbol{\tau}\right)}_{\dot{\mathcal{S}}_\mathrm{dissip}} \underbrace{- \boldsymbol{S}:\nabla \frac{\boldsymbol{f}_c}{\rho}}_{\dot{\mathcal{S}}_\mathrm{control}} \label{eq:strain_sq}
\end{align}
where the last terms in Eqs.~(\ref{eq:KE})–(\ref{eq:strain_sq}) represent the sink terms introduced by the flow control.

\begin{figure}%[hbt!]
\centering
\begin{minipage}[b]{1.0\textwidth}
\centering    
    \begin{subfigure}{0.325\textwidth}  
    \centering
    \includegraphics[width=0.98\textwidth]{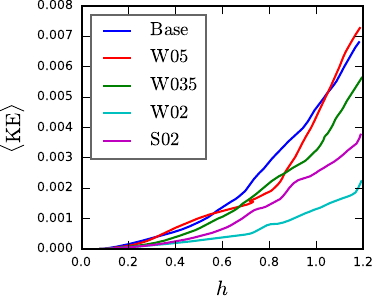}   
    \caption{}
    \end{subfigure} 
    \begin{subfigure}{0.325\textwidth}  
    \centering
    \includegraphics[width=0.98\textwidth]{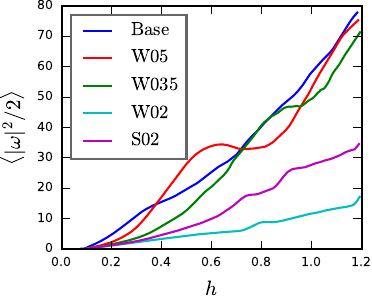}   
    \caption{}
    \end{subfigure} 
    \begin{subfigure}{0.325\textwidth}  
    \centering
    \includegraphics[width=0.98\textwidth]{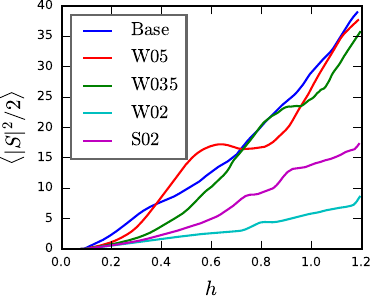}   
    \caption{}
    \end{subfigure} \\
    \begin{subfigure}{0.325\textwidth}  
    \centering
    \includegraphics[width=0.98\textwidth]{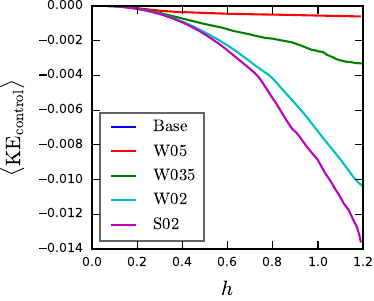}   
    \caption{}
    \end{subfigure} 
    \begin{subfigure}{0.325\textwidth}  
    \centering
    \includegraphics[width=0.98\textwidth]{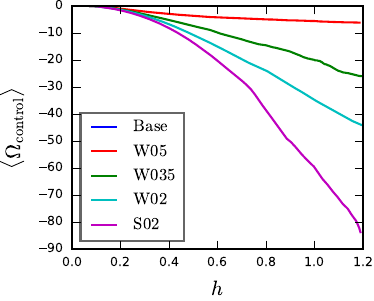}   
    \caption{}
    \end{subfigure} 
    \begin{subfigure}{0.325\textwidth}  
    \centering
    \includegraphics[width=0.98\textwidth]{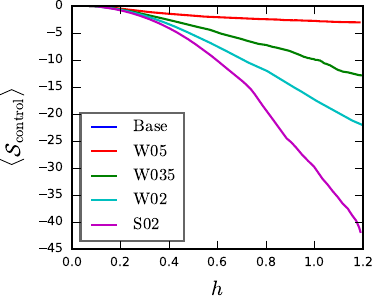}   
    \caption{}
    \end{subfigure}
\end{minipage}
    \caption{Evolution of mean kinetic energy (a), squared vorticity (b), and squared strain rate (c) as functions of the mixing width. Panels (d)–(f) show the corresponding control exerted on these three fields. \label{fig:KE_WS_evo}}
\end{figure}

Figures~\ref{fig:KE_WS_evo}(a)–(c) illustrate the evolution of mean kinetic energy (KE), mean enstrophy, and mean squared strain rate across various cases. The trends are consistent across the variables: they decrease monotonically with increasing control strength, following the order W05, W035, and W02. 
Comparing cases W02 and S02 reveals that preferential control over regions of high vorticity is more effective than control over regions of high strain rate in reducing both large-scale KE and small-scale vorticity and strain-rate. Additionally, the comparison between panels (b) and (c) indicates that the mean enstrophy evolution is approximately twice that of the mean squared strain rate. This aligns with the near-incompressible nature of the current RT simulations and is consistent with incompressible isotropic turbulence, where the purely kinematic relation $\langle |\boldsymbol{\omega}|^2 \rangle = 2\langle |\boldsymbol{S}|^2 \rangle$ holds.

To quantify the control strength, Figs.~\ref{fig:KE_WS_evo}(d)–(f) depict the magnitudes of the cumulative sink terms on the right-hand side of Eqs.~(\ref{eq:KE})–(\ref{eq:strain_sq}), defined as:  
\begin{align}
    \begin{split}
    \mathrm{KE}_\mathrm{control}(t) &= \int_0^t \dot{\mathrm{KE}}_\mathrm{control}(t') dt' \equiv -\int_0^t \bu\cdot \boldsymbol{f}_c~dt'; \\
    \mathrm{\Omega}_\mathrm{control}(t) &= \int_0^t \dot{\Omega}_\mathrm{control}(t') dt' \equiv -\int_0^t \boldsymbol{\omega}\cdot\nabla\times \frac{\boldsymbol{f}_c}{\rho}~dt'; \\
    \mathcal{S}_\mathrm{control}(t) &= \int_0^t \dot{\mathcal{S}}_\mathrm{control}(t') dt'\equiv -\int_0^t \boldsymbol{S}:\nabla\frac{\boldsymbol{f}_c}{\rho}~dt'
    \end{split}
\end{align}
Other cumulative budgets associated with Eqs.~(\ref{eq:KE})-(\ref{eq:strain_sq}) are defined in a similar way. The control terms shown in Figs.~\ref{fig:KE_WS_evo}(d)–(f) for KE, enstrophy, and squared strain rate exhibit a monotonic increase in magnitude as the control strength increases. When plotted as a function of the mixing width, the control magnitudes for KE, enstrophy, and squared strain-rate exhibit an approximately linear growth in the W05 and W035 cases. This behavior suggests that, in these cases, the instantaneous control magnitude saturates towards a fixed value at late times. However, in the W02 and S02 cases, the growth is faster than linear with respect to $h$, indicating highly nonlinear responses to the control parameter $p$ adopted in the control scheme, and their instantaneous control strength keeps growing without saturation. An unexpected behavior is observed in the larger control magnitude of case S02 compared to W02, even though the S02 case exhibits higher KE, enstrophy, and squared strain than W02. This indicates that preferential vorticity control is more effective than strain-rate control in mitigating nonlinear interactions among scales, resulting in reduced vortex stretching, slower energy cascade rate, and reduced turbulence intensity in the W02 case. This mechanism will be further elaborated upon in subsequent discussions.

\begin{figure}%[hbt!]
\centering
\begin{minipage}[b]{1.0\textwidth}
\centering    
    \begin{subfigure}{0.325\textwidth}  
    \centering
    \includegraphics[width=0.98\textwidth]{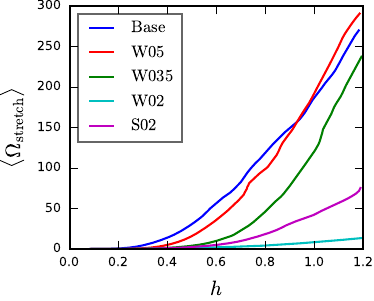}   
    \caption{}
    \end{subfigure} 
    \begin{subfigure}{0.325\textwidth}  
    \centering
    \includegraphics[width=0.98\textwidth]{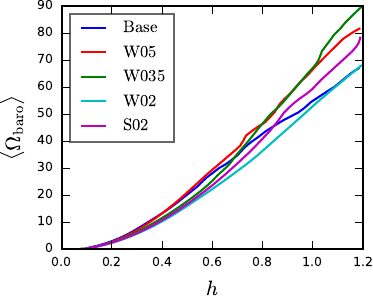}   
    \caption{}
    \end{subfigure} 
    \begin{subfigure}{0.325\textwidth}  
    \centering
    \includegraphics[width=0.98\textwidth]{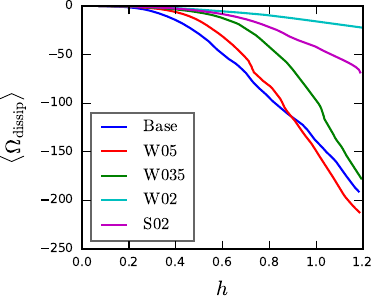}   
    \caption{}
    \end{subfigure} \\
    \begin{subfigure}{0.325\textwidth}  
    \centering
    \includegraphics[width=0.98\textwidth]{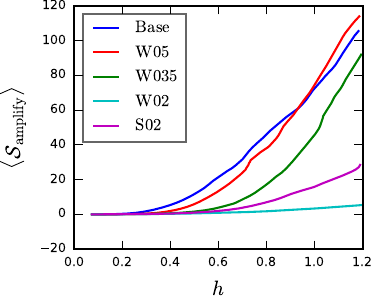}   
    \caption{}
    \end{subfigure} 
    \begin{subfigure}{0.325\textwidth}  
    \centering
    \includegraphics[width=0.98\textwidth]{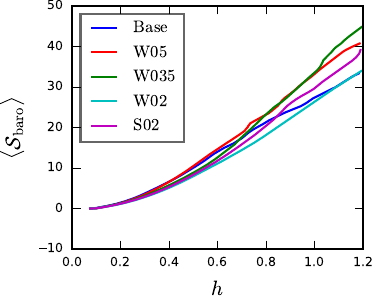}   
    \caption{}
    \end{subfigure} 
    \begin{subfigure}{0.325\textwidth}  
    \centering
    \includegraphics[width=0.98\textwidth]{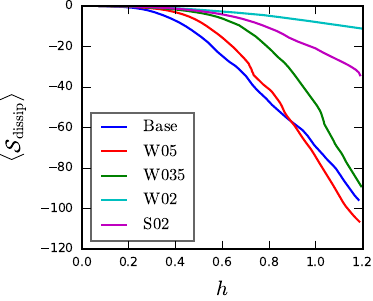}   
    \caption{}
    \end{subfigure}
\end{minipage}
    \caption{The mean cumulative budgets in enstrophy (panels (a)-(c)) and squared-strain (panels (d)-(f)) equations versus the mixing height.  \label{fig:S_W_budgets}}
\end{figure}

We present in Fig.~\ref{fig:S_W_budgets} the mean cumulative budgets of the terms in Eq.~(\ref{eq:vort_sq}) and (\ref{eq:strain_sq}) as functions of the mixing height, including contributions from vortex stretching, strain self-amplification, baroclinicity, and dissipation. Contributions from dilatation and pressure Hessian in current simulations are negligible and therefore omitted. Consistent with the behavior of vorticity and strain-rate, the mean values of the vorticity budgets share similar evolution trends with strain-rate budgets and are approximately twice in magnitude: $\langle \Omega_\mathrm{stretch}\rangle\approx 2\langle \mathcal{S}_\mathrm{amplify}\rangle$, $\langle \Omega_\mathrm{baro}\rangle\approx 2\langle \mathcal{S}_\mathrm{baro}\rangle$, and $\langle \Omega_\mathrm{dissip}\rangle\approx 2\langle \mathcal{S}_\mathrm{dissip}\rangle$ for all the cases reported here.

In conjunction with the control terms presented in Fig.~\ref{fig:KE_WS_evo}(e) and (f), we observe that vortex stretching, strain self-amplification, and baroclinic terms serve as sources of enstrophy and strain-rate, while dissipation and control terms act as sinks. The relative magnitudes of these terms underscore their differing contributions to vorticity and strain-rate, as well as their varying sensitivities to the control scheme.
Specifically, in the baseline case, vortex stretching and strain self-amplification dominate as the primary sources of vorticity and strain. In contrast, for the strongly controlled cases, the baroclinic terms emerge as the main contributors. 

The magnitude of vortex stretching and strain amplification is strongly influenced by the control strength. In weakly controlled cases (W05 and W035), their contributions show minimal reduction or even a slight increase compared to the baseline case. However, in strongly controlled cases (W02 and S02), their magnitudes decrease significantly.
In contrast, the baroclinic terms contribute only 1/3 to 1/2 as much as $\Omega_\mathrm{stretch}$ or $\mathcal{S}_\mathrm{amplify}$ in the baseline case. Nevertheless, their response to changes in control strength is relatively weak, with differences of up to 20\% between weakly and strongly controlled cases. Consequently, in cases W02 and S02, the baroclinic terms instead become the major source terms. This disparity arises because $\Omega_\mathrm{stretch}$ and $\mathcal{S}_\mathrm{amplify}$ are intrinsically linked to the energy cascade driven by turbulent interactions, which are more sensitive to the turbulence levels under different control strengths. In contrast, the baroclinic term is primarily governed by density stratification, which remains relatively consistent across different cases.
Regarding the sink terms, dissipation is significantly lower in strongly controlled cases due to the reduced energy cascade levels in these scenarios, a finding consistent with the evolution of RT coherent structures.

To further analyze the distribution of mass and kinetic energy across different length scales, we examine the spectra of mass fraction, velocity, and vorticity. In RT flows, the domain is inherently inhomogeneous and anisotropic, making it difficult to compute Fourier spectra along the non-periodic direction. To address this issue, a filtering spectrum approach is proposed in Sadek and Aluie \citep{SadekAluie18PRF}, which employs spatial coarse-graining to decompose the energy content of any flow variable into distinct scales. The filtering spectrum is defined as \citep{SadekAluie18PRF}:  
\begin{align}
    E(k_\ell) \equiv \frac{d}{dk_\ell} \left\langle \frac{1}{2} |\OL{u}_\ell(\bx)|^2 \right\rangle,
\end{align}
where $k_\ell = L_z/ \ell$  is the filtering wavenumber corresponding to the filtering width $\ell$, and $\OL{u}_\ell(\bx) = \int \mathrm{d}^3\br~ G_\ell(\bx - \br) u(\br)$ represents the coarse-graining (filtering) operation applied to the field variable $u(\bx)$. The angle bracket $\langle \cdot \rangle$ denotes spatial averaging, and $G_\ell(\br)$ is the filtering kernel with a characteristic width $\ell$. In this study, we adopt a Gaussian kernel, given by $G_\ell(\br) = \left(\frac{6}{\pi \ell^2}\right)^{3/2} e^{-6|\br|^2/\ell^2}$ \citep{Piomellietal91,John2003LES,Zhao18PRF}. Further extensions to the filtering spectrum have been developed to enhance its capabilities, such as using a high-pass filter to better resolve fields at small scales \citep{Sequential_filter} or characterizing the shape anisotropy of flow variables using multi-dimensional filters \citep{ZhaoAluie2023}.

\begin{figure}%[hbt!]
\centering
\begin{minipage}[b]{1.0\textwidth}
\centering    
    \begin{subfigure}{0.325\textwidth}  
    \centering
    \includegraphics[width=0.98\textwidth]{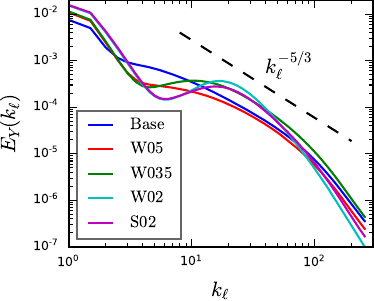}   
    \caption{}
    \end{subfigure} 
    \begin{subfigure}{0.325\textwidth}  
    \centering
    \includegraphics[width=0.98\textwidth]{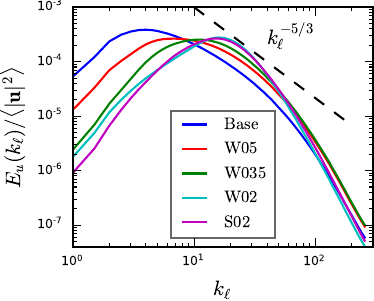}   
    \caption{}
    \end{subfigure} 
    \begin{subfigure}{0.325\textwidth}  
    \centering
    \includegraphics[width=0.98\textwidth]{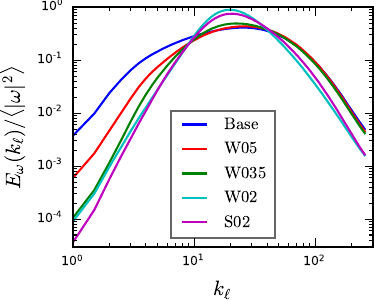}   
    \caption{}
    \end{subfigure}
\end{minipage}
    \caption{Filtering spectra of mass fraction (a), normalized velocity (b), and normalized vorticity (c) for various cases at the instant when the mixing width $h$ reaches 1.0. The filtering wavenumber associated with the filtering width $\ell$ is defined as $k_\ell=L_z/\ell$. \label{fig:filter_spec}}
\end{figure}

The filtering spectra results are presented in Fig.~\ref{fig:filter_spec}, illustrating the mass fraction, velocity, and vorticity spectra for various cases. In panel (a), the mass fraction spectrum highlights a distinctive feature at intermediate length scales: the two cases with strong control, W02 and S02, exhibit pronounced peaks around $k_\ell = 20$, corresponding to $\ell = L_x / 10$. This observation aligns with the persistent coherent bubble and spike structures visualized in W02 and S02 in Fig.~\ref{fig:RT_viz}, suggesting a strong influence of the control mechanisms on the intermediate-scale dynamics.
At large scales (small $k_\ell$), the mass fraction content is smallest in the baseline case and increases progressively with control strength. This behavior is attributed to the suppression of mass fraction flux in active vorticity or strain-rate controlled cases, which typically transfers the mass fraction variance from large to small scales \citep{ZhaoLi25}. The suppression effect slows the depletion of large-scale mass fraction content initially present in the density stratification. 
Conversely, at small scales (large $k_\ell$), the baseline case generally exhibits higher mass fraction content than the controlled cases, consistent with the same mechanism. This indicates the presence of more small-scale structures and, consequently, better mixing in the baseline case. However, an exception is observed for the W035 case, which shows greater small-scale mass fraction content than both the baseline and other controlled cases. This anomaly aligns with the mixedness value presented in Fig.~\ref{fig:width_theta}(c), where the W035 case demonstrates the highest mixing level among all cases.

The velocity and vorticity spectra are shown in Figs.~\ref{fig:filter_spec}(b) and (c). Unlike the mass fraction field, which lacks any external input, kinetic energy is supplied by the release of potential energy through gravity, while enstrophy is enhanced by the vortex stretching term. Consequently, significant differences in magnitude are evident for the velocity and vorticity spectra among the various cases. 
The baseline case exhibits the highest levels of kinetic energy and enstrophy spectra, with the magnitudes decreasing as the control strength increases. Furthermore, discrepancies exist in the length scales associated with the peak values of the velocity and vorticity spectra. In baseline RT turbulence, the peak of the KE spectrum gradually shifts to larger scales, broadening the inertial range as the RT mixing width grows \citep{Zhao22JFM}. In contrast, the controlled cases, particularly W02 and S02, show KE spectra peaking around $k_\ell = 20$, corresponding to the wavenumber associated with the mass fraction peaks. This alignment reflects the coherent bubble and spike structures characteristic of these flows. 
For the vorticity field, as shown in Fig.~\ref{fig:filter_spec}(c), the peak locations for the baseline and moderately controlled cases shift to much smaller scales compared to the KE spectrum peaks, indicating a separation between the large energy-containing scales (KE spectrum peaks) and the dissipation scales (vorticity spectrum peaks). However, in the heavily controlled cases W02 and S02, the peak locations of the KE and vorticity spectra are close, suggesting that no significant turbulence develops in these flows.

\subsection{Influence of flow control on the RT anisotropy}
\begin{figure}%[hbt!]
\centering
\begin{minipage}[b]{1.0\textwidth}
\centering 
    \begin{subfigure}{0.325\textwidth}  
    \centering
    \includegraphics[width=0.98\textwidth]{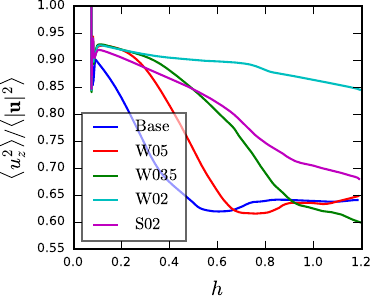}   
    \caption{}
    \end{subfigure} 
    \begin{subfigure}{0.325\textwidth}  
    \centering
    \includegraphics[width=0.98\textwidth]{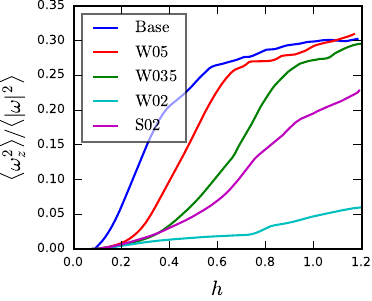}   
    \caption{}
    \end{subfigure}
    \begin{subfigure}{0.325\textwidth}  
    \centering
    \includegraphics[width=0.98\textwidth]{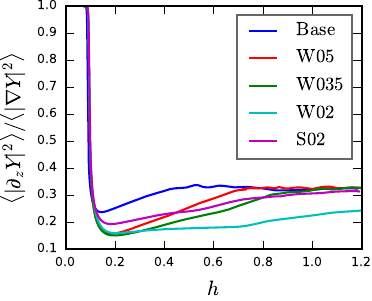}   
    \caption{}
    \end{subfigure} 
\end{minipage}
    \caption{Evolution of the ratio of the mean vertical component of (a) velocity, (b) vorticity, and (c) mass fraction gradient to their respective squared magnitudes as functions of the mixing width $h$. \label{fig:evo_aniso_ratio}}
\end{figure}

The adopted vorticity or strain-rate control significantly affects both the large- and small-scale structures in RT flows, thereby altering the anisotropy levels in these systems. Figure~\ref{fig:evo_aniso_ratio} illustrates the evolution of the ratio of the vertical components of velocity, vorticity, and mass fraction gradient to their respective squared magnitudes.  
At large scales, the velocity component ratio in Fig.~\ref{fig:evo_aniso_ratio}(a) asymptotes to approximately $\langle u_z^2\rangle / \langle|\bu|^2\rangle \approx 0.64$, corresponding to a large-scale anisotropy of $b_{zz} \equiv \langle u_z^2\rangle / \langle |\bu|^2\rangle - 1/3 \approx 0.307$. This value aligns with previously reported results for RT turbulence with $b_{zz} = 0.3$ \citep{RistorcelliClark04, Livescu09, Livescu10}.
In controlled cases, stronger control increases the vertical velocity component, as demonstrated by the W02 and S02 cases. Notably, in the W02 case, over 85\% of the kinetic energy is concentrated in the vertical component, indicating a substantial suppression of horizontal motion. This phenomenon arises because, in RT flows, horizontal kinetic energy is primarily generated by the pressure-strain term \citep{Zhou17-2}, $P (\partial_x u_x + \partial_y u_y)$. In the controlled cases, the reduced strain levels significantly deplete this term, thereby diminishing horizontal motion and enhancing vertical anisotropy.

In contrast, at small scales, as represented by vorticity or $\nabla Y$, the anisotropy in well-developed RT turbulence diminishes due to the chaotic interactions of small-scale structures \citep{CabotZhou13, Schneider16JFM}. This trend is evident in the baseline case, as shown in Figs.~\ref{fig:evo_aniso_ratio}(b) and (c). Here, both $\langle \omega_z^2 \rangle / \langle |\boldsymbol{\omega}|^2 \rangle$ and $\langle |\partial_z Y|^2 \rangle / \langle |\nabla Y|^2 \rangle$ approach 1/3 at late times, indicating that mean values of the three components become equal in magnitude.
However, in the controlled cases, where turbulence intensity is reduced or fully suppressed, the behavior of small-scale anisotropy deviates from that of the baseline case. Figs.~\ref{fig:evo_aniso_ratio}(b) and (c) confirm this difference, showing that the vertical components of vorticity and the mass fraction gradient are significantly smaller than the two horizontal components, especially in the W02 case. This anisotropy becomes more pronounced as the control magnitude increases.

\begin{figure}%[hbt!]
\centering
\begin{minipage}[b]{1.0\textwidth}
\centering 
    \begin{subfigure}{0.325\textwidth}  
    \centering
    \includegraphics[width=0.98\textwidth]{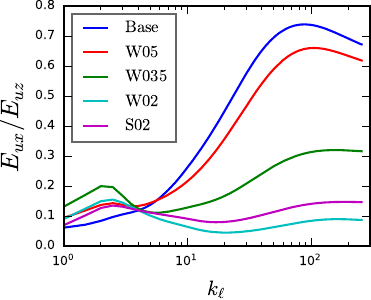}   
    \caption{}
    \end{subfigure} 
    \begin{subfigure}{0.325\textwidth}  
    \centering
    \includegraphics[width=0.98\textwidth]{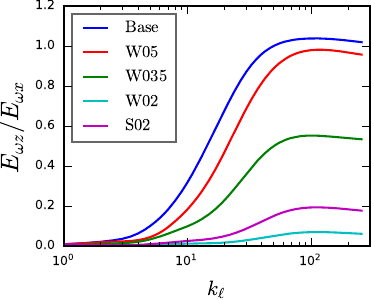}   
    \caption{}
    \end{subfigure}
    \begin{subfigure}{0.325\textwidth}  
    \centering
    \includegraphics[width=0.98\textwidth]{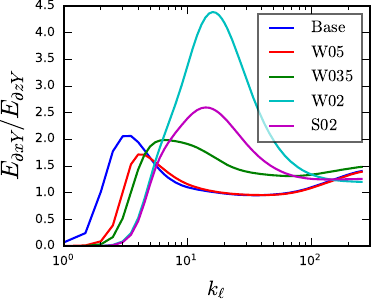}   
    \caption{}
    \end{subfigure} 
\end{minipage}
    \caption{Evolution of the spectra ratio between vertical and horizontal components: (a) ratio of horizontal to vertical velocity spectra, $E_{ux}/E_{uz}$, (b) ratio of vertical to horizontal vorticity spectra, $E_{\omega z}/E_{\omega x}$, and (c) ratio of horizontal to vertical mass fraction gradient spectra, $E_{\partial_x Y}/E_{\partial_z Y}$. All results are presented at the instant when the mixing width $h=1.0$. \label{fig:spec_aniso_ratio}}
\end{figure}

An alternative metric for directly quantifying anisotropy at different length scales is the ratio of vertical to horizontal spectra for density or velocity fields. Figure~\ref{fig:spec_aniso_ratio} illustrates this ratio for velocity, vorticity, and mass fraction gradients, providing insights into the scale-dependent anisotropic behavior of the flow.
In Fig.~\ref{fig:spec_aniso_ratio}(a), the ratio of horizontal to vertical velocity spectra reveals distinct trends across scales and control strengths. At large scales (small $k_\ell$), the horizontal velocity components are relatively weak in all cases. At small scales (large $k_\ell$), the ratio for the baseline case approaches 0.7–0.8, indicating near-isotropic behavior in the velocity field. However, as control strength increases, this ratio decreases, reflecting a suppression of horizontal motion at smaller scales and a consequent increase in anisotropy.
In Fig.~\ref{fig:spec_aniso_ratio}(b), the vertical-to-horizontal vorticity spectra ratio for the baseline case demonstrates that, at large scales, vertical vorticity is negligible compared to horizontal vorticity. At small scales, however, the spectra for vertical and horizontal vorticity become nearly equal, reflecting isotropy among the different vorticity components. This behavior is consistent with Kolmogorov's small-scale isotropy assumption. With increasing control strength, however, the ratio at small scales decreases significantly, leading to enhanced anisotropy in the vorticity field.
Together, Figs.~\ref{fig:spec_aniso_ratio}(a) and (b) highlight the critical role of small-scale vorticity and strain-rate fields in promoting isotropy in Rayleigh-Taylor (RT) flows. The suppression of these small-scale structures under stronger control conditions leads to a breakdown of isotropy, underscoring their importance in maintaining balanced anisotropic behavior in the flow.

The behavior of the mass fraction gradient, as shown in Fig.~\ref{fig:spec_aniso_ratio}(c), is more complex and exhibits distinct characteristics across different scales. At very large scales, comparable to the domain width ($k_\ell \leq 2$), the mass fraction gradient is dominated by the initial stratification, which varies predominantly in the vertical direction. This results in $E_{\partial_x Y} \ll E_{\partial_z Y}$, where the horizontal gradient is negligible compared to the vertical gradient.
At intermediate to small scales, however, the RT flow significantly alters this relationship. The horizontal gradient $\partial_x Y$ is strongly influenced by the coherent bubbles and spikes, which are aligned alternately in space. These structures produce large variations in $\partial_x Y$, reflecting the alternating upward and downward motion of heavy and light fluids. In contrast, the vertical gradient $\partial_z Y$ at intermediate to small scales is primarily driven by the horizontal motion generated by secondary Kelvin-Helmholtz instabilities. These instabilities cause the heavy and light fluids to overturn and move chaotically, contributing to the vertical gradient. The interplay of these two competing mechanisms results in $E_{\partial_x Y} \geq E_{\partial_z Y}$ at intermediate to small scales.

The anisotropy ratio, $E_{\partial_x Y} / E_{\partial_z Y}$, increases with control strength, as the suppression of horizontal motion becomes more pronounced. In the W02 case, this ratio reaches a peak value of approximately 4.5 at $k_\ell = 16$, corresponding to the characteristic width of the dominant bubbles and spikes. 
Interestingly, for the baseline case, the mass fraction gradient exhibits distinct anisotropic behavior across different scale regimes. At both large and small scales, the gradient is anisotropic, while at intermediate scales, it approaches near-isotropic behavior. This observation is consistent with the findings reported in Ref.~\cite{Livescu09}, where isotropy in RT flows is recovered only at intermediate scales. However, the application of flow control tends to disrupt this behavior, leading to a breakdown of the intermediate-scale isotropy.

\subsection{Small-scale alignment statistics}
In canonical turbulence, the vorticity vector exhibits a preferential alignment with the eigenvector corresponding to the intermediate eigenvalue of the strain-rate tensor, while the scalar gradient tends to align with the eigenvector associated with the smallest eigenvalue \cite{ashurst1987alignment}. This alignment behavior is also observed in RT turbulence at the incompressible limit \cite{CabotZhou13, livescu2021rayleigh}. These alignment trends are closely linked to the stretching dynamics of vortex and the scalar fields, thereby influencing the cascade rates of kinetic energy and scalar variance. In this section, we investigate the effects of flow control on these alignment statistics to assess its impact on the cascading processes of energy and scalar variance.

\begin{figure}%[hbt!]
\centering
\begin{minipage}[b]{1.0\textwidth}
\centering 
    \begin{subfigure}{0.47\textwidth}  
    \centering
    \includegraphics[width=0.98\textwidth]{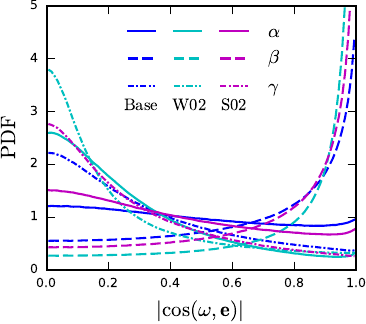}   
    \caption{}
    \end{subfigure} 
    \begin{subfigure}{0.49\textwidth}  
    \centering
    \includegraphics[width=0.98\textwidth]{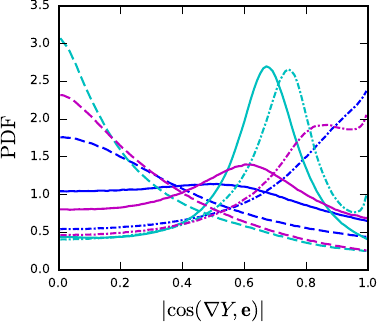}   
    \caption{}
    \end{subfigure}
\end{minipage}
    \caption{Alignment statistics between the eigenvectors of the strain-rate tensor and (a) the vorticity vector, as well as (b) the scalar gradient. The baseline, W02, and S02 cases are included.  \label{fig:alignment_stat}}
\end{figure}

Figure~\ref{fig:alignment_stat} illustrates the alignments between the strain-rate tensor, vorticity, and scalar gradient for three different cases. The vectors $e_\alpha$, $e_\beta$, and $e_\gamma$ respectively represent the eigenvectors associated with the eigenvalues in descending order of magnitude. Specifically, Fig.~\ref{fig:alignment_stat}(a) focuses on the alignment between the strain-rate tensor and vorticity. For all three cases, the vorticity shows a preferential alignment with the eigenvector corresponding to the intermediate eigenvalue of the strain-rate tensor ($\boldsymbol{\omega} \parallel \boldsymbol{e}_\beta$), consistent with trends observed in canonical turbulence. This preferential alignment is further enhanced under flow control, with the S02 and W02 cases exhibiting a stronger tendency towards this alignment. Additionally, Fig.~\ref{fig:alignment_stat}(b) examines the alignment of the scalar gradient with the strain-rate eigen-system. In the baseline case, the scalar gradient is predominantly aligned with the eigenvector associated with the smallest strain eigenvalue. However, when flow control is applied, this preferential alignment weakens in the S02 case and diminishes further in the W02 case. In the W02 case, the most probable alignment of the scalar gradient,  where $|\cos(\nabla Y, e_\alpha)|\approx |\cos(\nabla Y, e_\gamma)\approx 0.7$, lies in the plane spanned by the smallest ($\boldsymbol{e}_\gamma$) and largest ($\boldsymbol{e}_\alpha$) eigenvectors, forming an angle of approximately $45^\circ$ with both. It is important to note that a simple reduction in the Reynolds number does not result in such substantial alterations in the alignment statistics. This is evident from the comparison of the alignment between the baseline and the lowRe cases, as detailed in Appendix \ref{Appsec:align}.

To gain physical insights into the observed alignment statistics, we examine the evolution equations for vorticity and the scalar gradient, which are respectivly given by
\begin{align}
\frac{D}{Dt}\boldsymbol{\omega} &= \underbrace{(\bomega\cdot\nabla)\bu\vphantom{\frac{1}{\rho}}}_{\dot{\omega}_\mathrm{stretch}} \underbrace{-\bomega(\nabla\cdot\bu)\vphantom{\frac{1}{\rho}}}_{\dot{\omega}_\mathrm{dila}} + \underbrace{\frac{1}{\rho^2}(\nabla\rho\times\nabla P)}_{\dot{\omega}_\mathrm{baro}} + \underbrace{\nu\nabla^2\bomega\vphantom{\frac{1}{\rho}}}_{\dot{\omega}_\mathrm{visc}} \underbrace{-\nabla\times \frac{\boldsymbol{f}_c}{\rho}}_{\dot{\omega}_\mathrm{control}} \label{eq:vort_budget} \\
\frac{D}{Dt} \nabla Y &= \underbrace{-\frac{1}{2} (\nabla u +\nabla u^T)\cdot\nabla Y}_{\dot{\mathcal{N}}_\mathrm{stretch}} \underbrace{- \frac{1}{2}(\nabla u-\nabla u^T)\cdot\nabla Y}_{\dot{\mathcal{N}}_\mathrm{rot}} + \underbrace{\nabla(\frac{1}{\rho}\nabla\cdot(\rho D\nabla Y))}_{\dot{\mathcal{N}}_\mathrm{diffuse}} \label{eq:dY_budget}
\end{align}
The contributions to the changes in the alignment of vorticity or scalar gradient with respect to the strain eigenvectors arise from the combined effects of the terms on the right-hand sides (RHS) of Eqs.~(\ref{eq:vort_budget}) and (\ref{eq:dY_budget}). This is based on the assumption that the rotational rate of change of the strain eigenvector basis is a relatively slow process \cite{NomuraPost98JFM}. We will quantify the contributions of each term on the RHS of these equations to the differing alignment statistics due to different levels of flow control.

\begin{figure}%[hbt!]
\centering
\begin{minipage}[b]{1.0\textwidth}
\centering 
    \begin{subfigure}{0.43\textwidth}  
    \centering
    \includegraphics[width=0.98\textwidth]{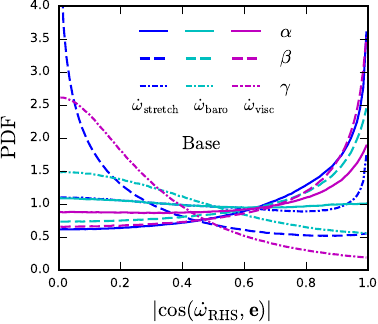}   
    \caption{}
    \end{subfigure} 
    \begin{subfigure}{0.43\textwidth}  
    \centering
    \includegraphics[width=0.98\textwidth]{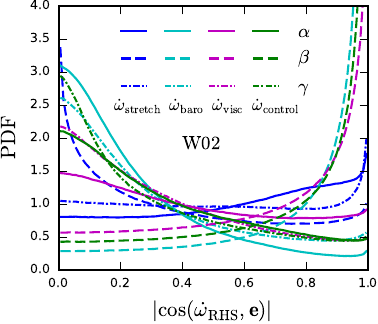}   
    \caption{}
    \end{subfigure} \\
    \begin{subfigure}{0.43\textwidth}  
    \centering
    \includegraphics[width=0.98\textwidth]{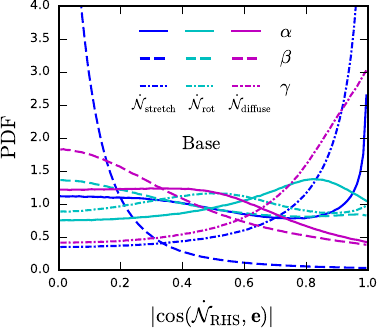}   
    \caption{}
    \end{subfigure} 
    \begin{subfigure}{0.43\textwidth}  
    \centering
    \includegraphics[width=0.98\textwidth]{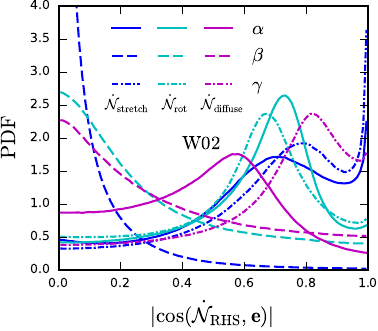}   
    \caption{}
    \end{subfigure}
\end{minipage}
    \caption{The alignment between the three strain rate eigenvectors and the terms on the RHS of Eq.~(\ref{eq:vort_budget}) (panels (a) and (b)) and Eq.~(\ref{eq:dY_budget}) (panels (c) and (d)). Panels (a) and (c) correspond to the baseline case, whereas panels (b) and (d) correspond to the W02 case. \label{fig:alignment_stat_RHS}}
\end{figure}

Figure~\ref{fig:alignment_stat_RHS} compares the alignment of the individual terms on the RHS of Eqs.~(\ref{eq:vort_budget}) and (\ref{eq:dY_budget}) with the strain-rate eigenvectors. In panel (a), the vorticity budget for the baseline case is presented, showing the contributions from the $\dot{\omega}_\mathrm{stretch}$, $\dot{\omega}_\mathrm{baro}$, and $\dot{\omega}_\mathrm{visc}$ terms. In panel (b), which corresponds to the flow-controlled case, the additional contribution from the $\dot{\omega}_\mathrm{control}$ term is included for comparison.
In the baseline case (panel a), the $\dot{\omega}_\mathrm{stretch}$ term primarily aligns with the largest strain-rate eigenvector $e_\alpha$, while the $\dot{\omega}_\mathrm{baro}$ and $\dot{\omega}_\mathrm{visc}$ terms exhibit a stronger alignment with the intermediate strain-rate eigenvector $e_\beta$.. Conversely, the baroclinic, viscous, and control terms enhance the alignment of vorticity with $e_\beta$.
A similar trend is observed in the W02 flow-controlled case (panel b), with the inclusion of the $\dot{\omega}_\mathrm{control}$ term further reinforcing the alignment with $e_\beta$. This suggests that, while vortex stretching weakens the alignment of vorticity with $e_\beta$ by promoting more efficient stretching along $e_\alpha$, the baroclinic, viscous, and control terms collectively enhance this alignment. The interplay between these terms highlights the complex dynamics governing vorticity evolution and its relationship with strain-rate eigenvectors in both baseline and controlled flows.

\begin{figure}%[hbt!]
\centering
\begin{minipage}[b]{1.0\textwidth}
\centering 
    \begin{subfigure}{0.43\textwidth}  
    \centering
    \includegraphics[width=0.98\textwidth]{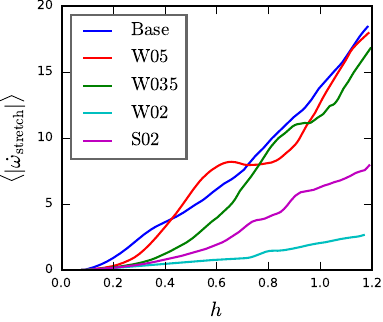}   
    \caption{}
    \end{subfigure} 
    \begin{subfigure}{0.43\textwidth}  
    \centering
    \includegraphics[width=0.98\textwidth]{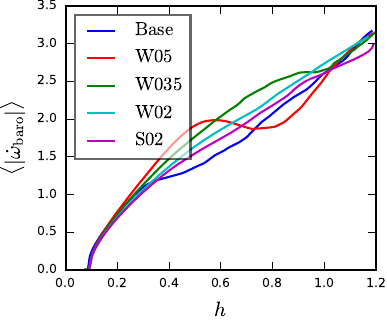}   
    \caption{}
    \end{subfigure} \\
    \begin{subfigure}{0.43\textwidth}  
    \centering
    \includegraphics[width=0.98\textwidth]{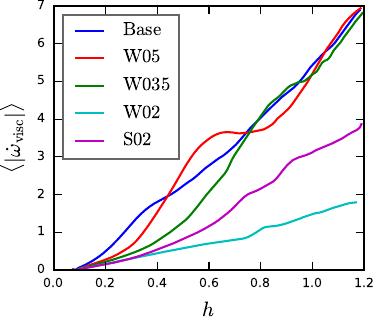}   
    \caption{}
    \end{subfigure} 
    \begin{subfigure}{0.43\textwidth}  
    \centering
    \includegraphics[width=0.98\textwidth]{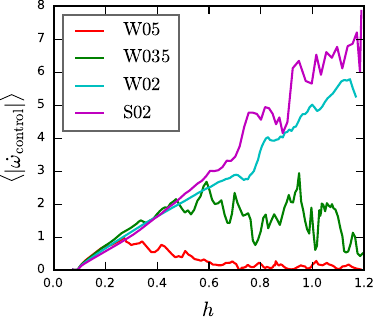}   
    \caption{}
    \end{subfigure}
\end{minipage}
    \caption{The temporal evolution of the average mean magnitude of the vorticity budgets of Eq.~(\ref{eq:vort_budget}) with respect to the mixing width. Panels (a)-(d) illustrate the evolution of the magnitudes corresponding to the $\dot{\omega}_\mathrm{stretch}, \dot{\omega}_\mathrm{baro}, \dot{\omega}_\mathrm{visc}, \dot{\omega}_\mathrm{control}$ terms, respectively. \label{fig:mag_W_RHS}}
\end{figure}

Figure~\ref{fig:mag_W_RHS} further emphasizes that, in the baseline case, the $\dot{\omega}_\mathrm{stretch}$ and $\dot{\omega}_\mathrm{visc}$ terms dominate the other terms in magnitude. In contrast, for the W02 flow-controlled case, all four terms on the RHS of Eq.~(\ref{eq:vort_budget}) exhibit comparable magnitudes. 
This difference in the relative contributions of the terms leads to distinct alignment behaviors between the vorticity vector $\bomega$ and the strain-rate eigenvector $e_\beta$. Specifically, the combined effect of the vorticity budget in the controlled case results in a stronger alignment between $\bomega$ and $e_\beta$ compared to the baseline case, as illustrated in Fig.~\ref{fig:alignment_stat}(a). 

\begin{figure}%[hbt!]
\centering
\begin{minipage}[b]{1.0\textwidth}
\centering 
    \begin{subfigure}{0.325\textwidth}  
    \centering
    \includegraphics[width=0.98\textwidth]{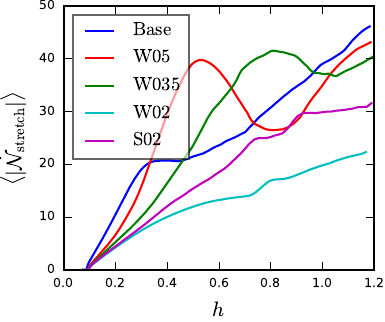}   
    \caption{}
    \end{subfigure} 
    \begin{subfigure}{0.325\textwidth}  
    \centering
    \includegraphics[width=0.98\textwidth]{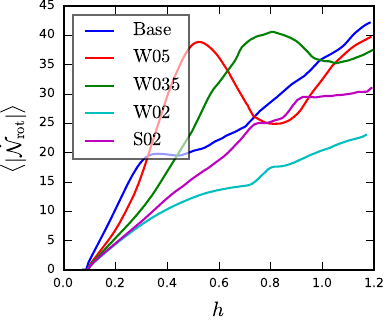}   
    \caption{}
    \end{subfigure}
    \begin{subfigure}{0.325\textwidth}  
    \centering
    \includegraphics[width=0.98\textwidth]{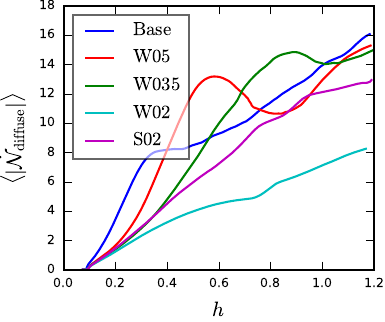}   
    \caption{}
    \end{subfigure} 
\end{minipage}
    \caption{The temporal evolution of the average mean magnitude of the scalar gradient budgets of Eq.~(\ref{eq:dY_budget}) with respect to the mixing width. Panels (a)-(c) show the evolution of the magnitudes of $\dot{\mathcal{N}}_\mathrm{stretch}, \dot{\mathcal{N}}_\mathrm{rot}, \dot{\mathcal{N}}_\mathrm{diffuse}$, respectively. \label{fig:mag_dY_RHS}}
\end{figure}

In Fig.~\ref{fig:alignment_stat_RHS} (c), the alignment of the scalar gradient budgets for the baseline case reveals that the stretching term ($\dot{\mathcal{N}}_\mathrm{stretch}$) and the diffusion term ($\dot{\mathcal{N}}_\mathrm{diffuse}$) primarily align with the smallest eigenvector $e_\gamma$. In contrast, the rotational term ($\dot{\mathcal{N}}_\mathrm{rot}$) shows no preferential alignment with any of the three eigenvectors. 
For the W02 case in Fig.~\ref{fig:alignment_stat_RHS} (d), the stretching term ($\dot{\mathcal{N}}_\mathrm{stretch}$) also tends to align with $e_\gamma$, but this tendency is reduced, resulting in an enhanced likelihood of alignment with $e_\alpha$. Additionally, the rotational term ($\dot{\mathcal{N}}_\mathrm{rot}$) exhibits a $45^{\circ}$ alignment with both $e_\alpha$ and $e_\gamma$. The magnitudes of the three terms ($\dot{\mathcal{N}}_\mathrm{stretch}$, $\dot{\mathcal{N}}_\mathrm{rot}$, and $\dot{\mathcal{N}}_\mathrm{diffuse}$) are comparable, as is illustrated in Fig.~\ref{fig:mag_dY_RHS}.
The combined effect of these alignments results in the final configuration shown in Fig.~\ref{fig:alignment_stat} (b) for the W02 case, where $\nabla Y$ is perpendicular to $e_\beta$ but forms a $45^{\circ}$ angle with both $e_\alpha$ and $e_\gamma$.

\subsection{Dynamics governing different control levels}
In this section, we perform a detailed analysis of the vorticity and strain-rate statistics, with a particular focus on elucidating the dynamical effects of control strength on the spatial distribution of the small-scale quantities. Our previous observations suggest the existence of a critical control parameter, $p$, below which RT turbulence is heavily suppressed, specifically at $p = 0.2$. We will explore the underlying reasons for this suppression and discuss the implications of varying levels of vorticity control on the overall flow dynamics.

\begin{figure}%[hbt!]
\centering
\begin{minipage}[b]{1.0\textwidth}
\centering    
    \begin{subfigure}{0.4\textwidth}  
    \centering
    \includegraphics[width=0.98\textwidth]{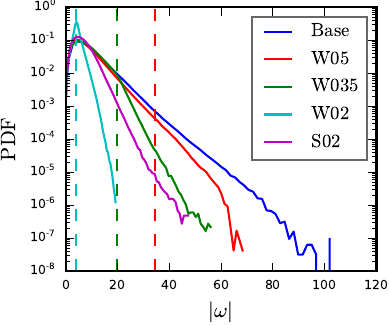}   
    \caption{}
    \end{subfigure} 
    \begin{subfigure}{0.4\textwidth}  
    \centering
    \includegraphics[width=0.98\textwidth]{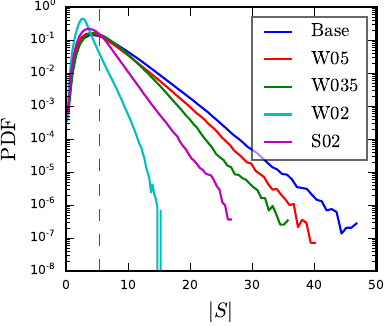}   
    \caption{}
    \end{subfigure} \\
    \begin{subfigure}{0.4\textwidth}  
    \centering
    \includegraphics[width=0.98\textwidth]{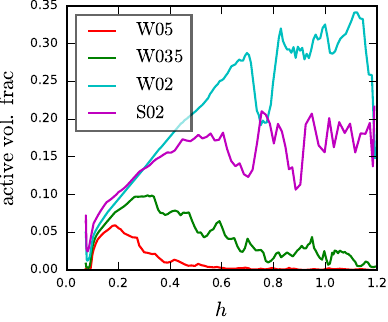}   
    \caption{}
    \end{subfigure} 
    \begin{subfigure}{0.4\textwidth}  
    \centering
    \includegraphics[width=0.98\textwidth]{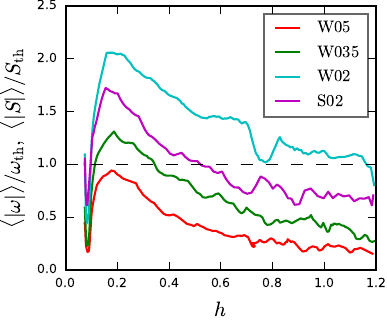}   
    \caption{}
    \end{subfigure} 
\end{minipage}
    \caption{The PDFs of the vorticity magnitude (a) and the strain-rate magnitude (b) for different simulation cases at the instant when the mixing width $h=1.0$. The dashed vertical lines in panels (a) and (b) represent the control thresholds $\omega_p$ and $S_p$ for the respective cases. (c) The volume fraction of the domain where the vorticity or strain-rate control is active as functions of the mixing width. (d) The ratio of mean vorticity (cases W05, W035, and W02) or mean strain-rate magnitude (case S02) to their respective control threshold, as functions of the mixing width $h$.\label{fig:PDF_SW_thresh}}
\end{figure}

Figure~\ref{fig:PDF_SW_thresh}(a) and (b) present the probability density functions (PDFs) of vorticity and strain-rate magnitudes for various cases at $h = 1.0$. As anticipated, the PDFs for the baseline RT case are broader compared to those of the controlled cases, with the width of the distributions decreasing as the control strength increases. This narrowing reflects the progressively stronger suppression of extreme fluctuations in vorticity and strain-rate magnitudes under higher levels of control.
The dashed vertical lines in Fig.~\ref{fig:PDF_SW_thresh}(a) and (b) denote the control thresholds $\omega_p \equiv p \, \mathrm{max}|\bomega|$ and $S_p \equiv p \, \mathrm{max}|S|$, and the regions where vorticity or strain-rate magnitudes exceed these thresholds are subject to control. In the W02 case, the control threshold lies below the mean value of the PDF, implying that vorticity control is applied across a larger portion of the domain. In contrast, for cases with weaker control, the control threshold is significantly above the mean value, limiting the spatial extent of control.

The evolution of the active volume fraction, where control is applied, is illustrated in Fig.~\ref{fig:PDF_SW_thresh}(c). For the W05 and W035 cases, the volume fraction initially increases with the mixing width but eventually decreases to zero, reaching a maximum value below 0.1. This behavior suggests a scale separation between the mean and maximum values of vorticity or strain-rate magnitudes, leading to the fat tails observed in the PDFs for the two cases.
In contrast, for the S02 and W02 cases, the controlled volume fractions remain constant or even increase over time. This indicates that the maximum growth rate of vorticity or strain-rate magnitudes in these cases is slower than the growth rate of their mean values. As a result, the suppression of extreme fluctuations is more effective, thereby hindering the growth of turbulence and leading to narrower PDFs. These findings highlight the critical role of control thresholds and active volume fractions in modulating the statistical properties of vorticity and strain-rate fields, as well as their influence on turbulence dynamics.

Additionally, Fig.~\ref{fig:PDF_SW_thresh}(d) plots the ratio of the mean vorticity or strain-rate to the control threshold. For the W05 and W035 cases, this ratio remains mostly below 1, while for the S02 case, it is consistently above 1, indicating that critical control is achieved when the control threshold falls below the mean value.

The joint PDF of the second and third invariants of the velocity gradient tensor, $Q \equiv -\frac{1}{2} \partial_j u_i \partial_i u_j$ and $R \equiv -\frac{1}{3} \partial_j u_i \partial_k u_j \partial_i u_k$, provides valuable insights into the small-scale flow topology. These invariants are central to understanding the dynamics of turbulence, as $Q$ represents the difference between squared vorticity and squared strain-rate, while $R$ characterizes the competition between strain self-amplification and vortex stretching. The joint PDF of $Q$ and $R$ exhibits the characteristic teardrop shape, a feature commonly observed in incompressible turbulence \cite{Meneveau2011, JohnsonWilczek2024} and in RT turbulence under the incompressible limit \cite{ZhaoLi25}. This teardrop-shaped distribution serves as a hallmark of the underlying dynamical processes that govern small-scale structures in turbulent flows, offering a quantitative framework for characterizing the intricate interplay between vorticity and strain-rate dynamics.
In fully developed turbulence, the normalized joint PDFs of $Q^* \equiv Q/\langle \omega^2\rangle$ and $R^* \equiv R/\langle \omega^2\rangle^{3/2}$ exhibit similar magnitudes for the same type of flow, provided they share comparable energy transfer pathways \cite{JohnsonWilczek2024}. 

\begin{figure}%[hbt!]
\centering
\begin{minipage}[b]{1.0\textwidth}
\centering    
    \begin{subfigure}{0.4\textwidth}  
    \centering
    \includegraphics[width=0.98\textwidth]{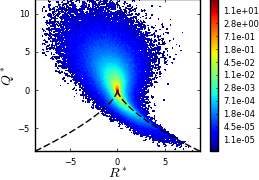}   
    \caption{$\mathrm{Base}$}
    \end{subfigure} 
    \begin{subfigure}{0.4\textwidth}  
    \centering
    \includegraphics[width=0.98\textwidth]{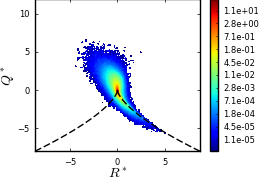}   
    \caption{$\mathrm{W035}$}
    \end{subfigure} \\
    \begin{subfigure}{0.4\textwidth}  
    \centering
    \includegraphics[width=0.98\textwidth]{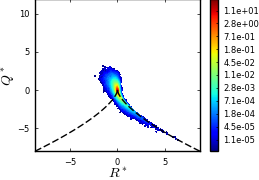}   
    \caption{$\mathrm{W02}$}
    \end{subfigure} 
    \begin{subfigure}{0.4\textwidth}  
    \centering
    \includegraphics[width=0.98\textwidth]{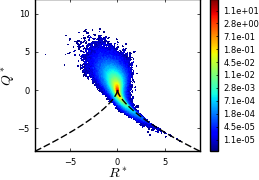}   
    \caption{$\mathrm{S02}$}
    \end{subfigure} 
\end{minipage}
    \caption{The joint PDFs of the normalized second and third velocity gradient invariants $Q^*$  and $R^*$ at a mixing width of $h=1.0$, where $Q^*\equiv Q/\langle\omega^2\rangle$ and $R^*\equiv R/\langle \omega^2\rangle^{3/2}$, for the four different cases shown in panels (a)-(d). \label{fig:jPDF_QR}}
\end{figure}

The effect of RT flow control on the joint PDFs between $Q$ and $R$ is illustrated in Fig.~\ref{fig:jPDF_QR}. In panel (a), the baseline case retains the classical teardrop shape, with pronounced tails in the quadrants $Q<0, R>0$ and $Q>0, R<0$. Under flow control, the teardrop shape contracts toward the origin and the right Vieillefosse tail (right dashed line), indicating a significant reduction in vorticity, strain magnitude, strain self-amplification, and vortex stretching. This trend aligns with the relative magnitudes of the stretching and strain amplification budgets $\dot{\Omega}_\mathrm{stretch}$ and $\dot{\mathcal{S}}_\mathrm{amplify}$ shown in fig.~\ref{fig:S_W_budgets}, which are linked to turbulence intensity. 
Moreover, along the right Vieillefosse tail, which is primarily governed by shear-dominated motion, the maximum magnitude of $Q$ and $R$ on the normalized joint PDFs remains similar even between the baseline and the strongly controlled W02 case. This highlights the persistence of high-shear regions in RT turbulence irrespective of flow control.

\begin{figure}%[hbt!]
\centering
\begin{minipage}[b]{1.0\textwidth}
\centering    
    \begin{subfigure}{0.4\textwidth}  
    \centering
    \includegraphics[width=0.98\textwidth]{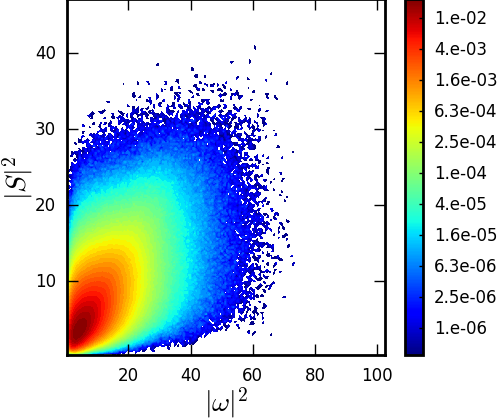}   
    \caption{$\mathrm{Base}$}
    \end{subfigure} 
    \begin{subfigure}{0.4\textwidth}  
    \centering
    \includegraphics[width=0.98\textwidth]{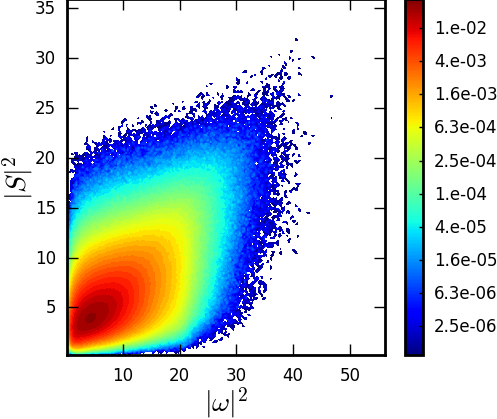}   
    \caption{$\mathrm{W035}$}
    \end{subfigure} \\
    \begin{subfigure}{0.4\textwidth}  
    \centering
    \includegraphics[width=0.98\textwidth]{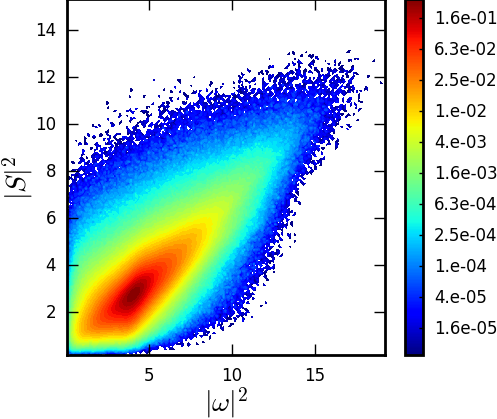}   
    \caption{$\mathrm{W02}$}
    \end{subfigure} 
    \begin{subfigure}{0.4\textwidth}  
    \centering
    \includegraphics[width=0.98\textwidth]{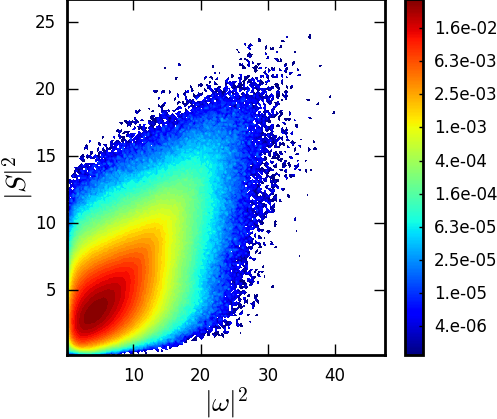}   
    \caption{$\mathrm{S02}$}
    \end{subfigure} 
\end{minipage}
    \caption{The joint PDFs between the squared magnitudes of strain-rate and vorticity fields for four cases in panels (a)-(d) at a mixing width of $h=1.0$. \label{fig:jPDF_SW}}
\end{figure}

The direct comparison between squared vorticity and strain-rate fields is illustrated in their joint PDFs shown in Fig.~\ref{fig:jPDF_SW} at $h = 1.0$. Across the baseline case and the controlled cases W035, W02, and S02, we observe not only differences in the magnitude of the joint PDFs but also an evolution in their shape. 
In the baseline RT case with developed turbulence, the joint PDFs exhibit a broader, more blunted shape. This reflects the complex and chaotic interplay between vorticity and strain-rate fields in a turbulent environment. In contrast, in the W02 case, where turbulence is significantly suppressed, the joint PDFs display a sharper profile. This transition in shape is closely correlated with the increase in control strength.
As a result of this evolution, regions of simultaneous high vorticity and high strain become more prevalent in the controlled cases. This suggests that the control mechanisms effectively suppress the extreme fluctuations typically associated with fully developed turbulence, leading to a tighter coupling between vorticity and strain-rate fields. In contrast, in the turbulent RT case, the regions of peak strain tend to be spatially separated from those of peak vorticity, reflecting the more disordered and chaotic evolution of these two quantities in a turbulent flow. 

\begin{figure}%[hbt!]
\centering
\begin{minipage}[b]{1.0\textwidth}
\centering    
    \begin{subfigure}{0.31\textwidth}  
    \centering
    \includegraphics[width=0.98\textwidth]{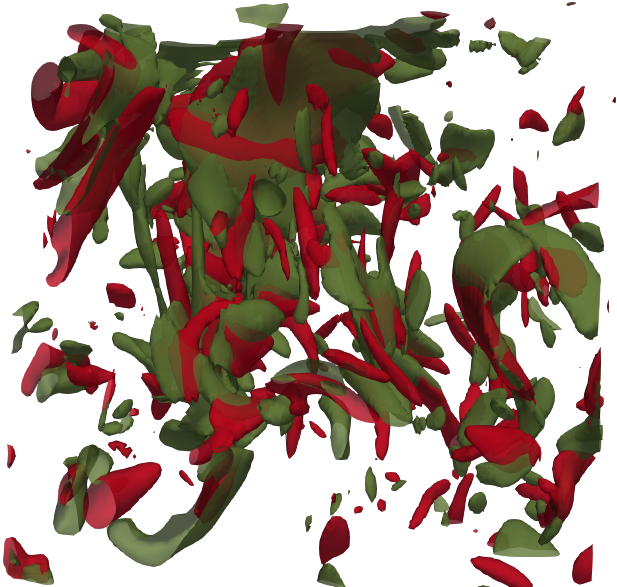}   
    \caption{$\mathrm{Base}$}
    \end{subfigure} 
    \begin{subfigure}{0.31\textwidth}  
    \centering
    \includegraphics[width=0.98\textwidth]{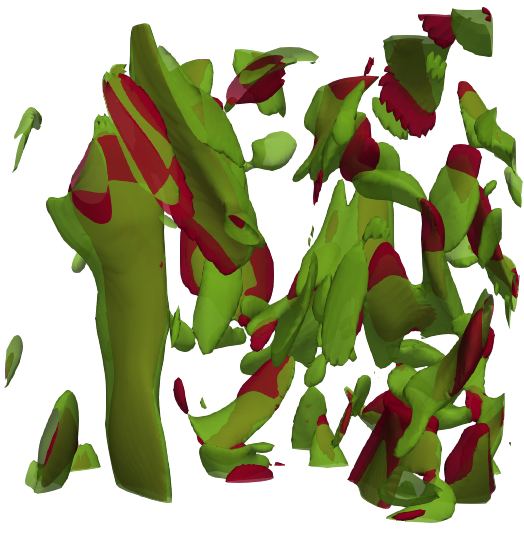}   
    \caption{$\mathrm{W02}$}
    \end{subfigure} 
    \begin{subfigure}{0.365\textwidth}  
    \centering
    \includegraphics[width=0.98\textwidth]{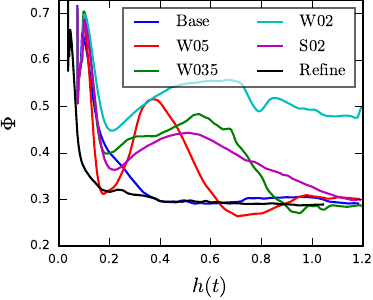}   
    \caption{volume ratio of two sets}
    \end{subfigure} 
\end{minipage}
    \caption{(a) and (b) Visualizations of the vorticity (red) and strain-rate (green) isosurfaces within a small region of the baseline and W02 cases, respectively. (c) The ratio of the volume occupied by regions with both high strain-rate and high vorticity to the volume occupied by regions where either the vorticity or strain-rate field is large. \label{fig:viz_SW}}
\end{figure}

The physical mechanisms underlying the vorticity-strain joint PDFs are further elucidated in Fig.~\ref{fig:viz_SW} (a) and (b), which provide visualizations of vorticity and strain-rate isosurfaces within a localized region of the simulation domain. In the turbulent baseline case (panel a), the red vorticity isosurfaces exhibit elongated, randomly oriented, worm-like structures, characteristic of the chaotic and disordered nature of turbulence. In contrast, the green strain-rate isosurfaces appear as sheet-like structures that surround the vorticity isosurfaces but remain spatially distinct, reflecting the decoupled evolution of strain and vorticity in fully developed turbulence.
In the W02 case (panel b), the turbulence is significantly suppressed, resulting in more regular and organized structures. Here, the vorticity isosurfaces are largely enclosed by the strain-rate isosurfaces, indicating that regions of high vorticity and high strain-rate are spatially coincident. These structures primarily align vertically, reflecting the shearing motion between rising bubbles and sinking spikes. This visualization highlights the transition from the chaotic, disordered dynamics of turbulence to a more structured and coherent flow regime under the influence of control.

To quantitatively analyze this observation, Fig.~\ref{fig:viz_SW}(c) presents the temporal evolution of the ratio $\Phi$, which is defined as the volume fraction of regions where both high strain-rate and high vorticity occur, relative to the volume where either high strain-rate or high vorticity is present. Mathematically, this is expressed as:
\begin{align*}
\Phi = \frac{\#\left\{\left(S > \langle S\rangle + \sigma_S\right) \cap \left(\omega > \langle \omega\rangle + \sigma_\omega\right)\right\}}{\#\left\{\left(S > \langle S\rangle + \sigma_S\right) \cup \left(\omega > \langle \omega\rangle + \sigma_\omega\right)\right\}},
\end{align*}
where $\#$ denotes the volume of a set, and $\sigma_S$ and $\sigma_\omega$ represent the root-mean-square values of the strain-rate and vorticity fields, respectively. 

The results reveal distinct trends across the cases. In the baseline case, $\Phi$ converges to a value of approximately 0.3, reflecting the relatively low likelihood of regions simultaneously exhibiting high strain-rate and high vorticity in a turbulent flow. In contrast, the weakly controlled cases (W05, W035, and S02) exhibit delayed convergence, while the heavily controlled case (W02) stabilizes at a significantly higher value of $\Phi \approx 0.5$, indicating a stronger tendency for high strain-rate and high vorticity regions to overlap.

This behavior indicates that in flows with lower turbulence levels, such as those in the controlled cases, the dominant shearing motion promotes the formation of large, coherent regions where both high strain-rate and high vorticity are present. In contrast, in highly turbulent flows, such as the baseline case, the self- and mutual-interactions between strain-rate and vorticity fields lead to the development of more random, small-scale structures. These interactions disrupt the alignment of high strain-rate and high vorticity regions, making their simultaneous occurrence less common and resulting in a lower value of $\Phi$.

\section{Conclusions}
\label{sec:conclusions}

In this paper, we have investigated the role of small-scale structures in turbulent Rayleigh-Taylor flows by applying preferential flow control to high vorticity or high strain-rate regions and performing detailed numerical simulations. Our results provide significant insights into the effects of flow control on turbulence dynamics, flow anisotropy, and mixing behavior.

The findings reveal that a high level of flow control promotes more regular bubble and spike structures, reduces mixedness, and enhances flow anisotropy. Flow control actively augments the alignment of vorticity and the eigenvector associated with the intermediate eigenvalue, while simultaneously weakening the alignment between the scalar gradient and the eigenvector of the smallest eigenvalue. The control mechanism leads to a more regular bubble and spike structures and thus resulting in the formation of simultaneous high vorticity and high strain-rate regions that enhance the coherence of the flow.

Specifically, flow anisotropy is significantly affected by flow control. In strongly controlled cases, the heavy-light fluid interface tends to align vertically rather than horizontally, and anisotropy persists across all scales, with the flow predominantly oriented in the vertical direction. In contrast, for the baseline case, the velocity and vorticity fields exhibit large-scale anisotropy with near-isotropy at small scales, while the scalar gradient shows isotropy only at intermediate scales. Flow control thus alters the scale-dependent anisotropy, enforcing a more directional flow with persistent anisotropy at all scales.

The alignment statistics of vorticity and strain fields are primarily governed by the interplay of vortex stretching, baroclinic, and viscous effects. Vortex stretching weakens the alignment between vorticity and the intermediate eigenvector, while baroclinic and viscous effects enhance it. Flow control suppresses the vortex stretching and viscous contributions, allowing vorticity and the intermediate eigenvector to align more closely. In contrast, for the scalar gradient, the stretching and rotation effects weaken its alignment with the smallest eigenvector, resulting in a less efficient downscale scalar variance flux in controlled cases.

Finally, the effect of flow control plays a crucial role in reducing regions of excess vorticity and strain-rate. By suppressing these extreme fluctuations, flow control leads to the formation of overlapped regions of high vorticity and high strain-rate, resulting in more organized and less turbulent flow structures. This suppression of turbulence intensity enhances the overall coherence of the flow. When the control parameter $p$ is lowered to apply control from the largest vorticity/strain-rate magnitudes down to their mean, RT turbulence is significantly suppressed. The investigation of flow control and small-scale structures on RT evolution could benefit realistic cases such as RT flow under magnetic fields and rotation, which all suppress small scale motions. Moreover, similar small-scale flow patterns may also emerge in other flow control scenarios, such as turbulence reduction by polymer additives \citep{White08ARFM,Procaccia2008RMP, Xi2019PoF} and heat transfer enhancement as well as flow relaminarization in turbulent Rayleigh-Bénard convection through external wall shear \citep{XuXi23JFM}. Applying similar analysis in these contexts could potentially yield valuable physical insights.

The study of flow control and its influence on small-scale structures in RT evolution holds considerable potential for advancing our understanding of realistic scenarios. For instance, RT flows under the influence of magnetic fields or rotation exhibit suppressed small-scale motions due to their stabilizing effects.  Findings presented here could provide valuable guidance for applications in such scenarios, which find applications in fusion plasmas, astrophysical phenomena, and other practical scenarios where turbulence suppression and flow coherence are of critical importance.

% \clearpage
\appendix

% \section{Appendix} 

\renewcommand{\theequation}{A-\arabic{equation}}
\renewcommand{\thefigure}{A\arabic{figure}}
\setcounter{equation}{0}  % reset counter 
\setcounter{figure}{0}  % reset counter 

\section{Visualizations of the Refine and the lowRe cases} \label{Appsec:viz}
\begin{figure}%[hbt!]
\centering
\begin{minipage}[b]{1.0\textwidth}
\centering    
    \begin{subfigure}{0.92\textwidth}  
    \centering
    \includegraphics[width=0.98\textwidth]{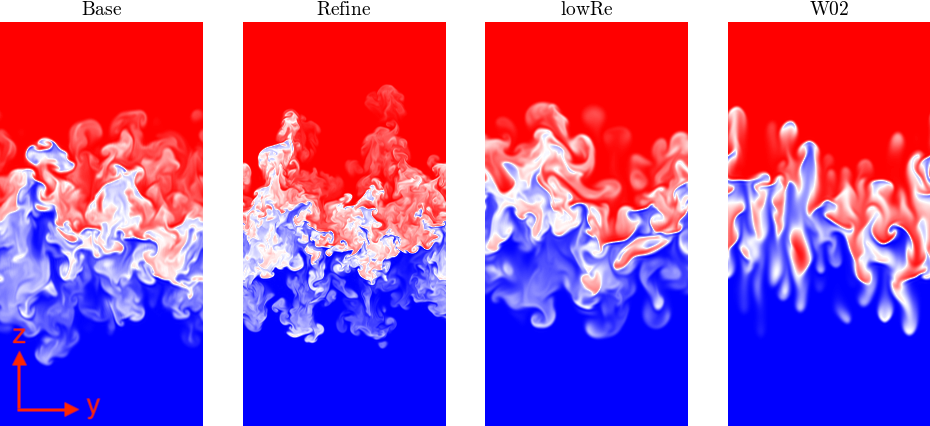}   
    \end{subfigure} 
\end{minipage}
    \caption{Visualizations of the mass fraction field at $y$-$z$ slices for the Base, Refine, lowRe, and W02 cases at the instance when the mixing width $h = 1.0$, similar to fig.~\ref{fig:RT_viz}. \label{Appfig:viz_refine_lowRe}}
\end{figure}

To complement the visualizations of the mass fraction fields presented in Fig.~\ref{fig:RT_viz}, the corresponding results for the Refine and lowRe cases are provided in Fig.~\ref{Appfig:viz_refine_lowRe}. These additional results reveal distinct characteristics of the flow in each case. Specifically, the refined case demonstrates a greater abundance of small-scale structures, indicative of a more resolved and detailed flow field. In contrast, the lowRe case exhibits a more smeared appearance, suggesting a coarser representation of the flow features. 
Despite the lower Reynolds number in the lowRe case compared to the W02 case, a notable observation is the presence of more complex and pronounced lateral motions in the lowRe case. This finding implies that turbulence suppression achieved through a uniform increase in viscosity, as in the lowRe case, is less effective than preferential flow control strategies that specifically target high vorticity or high strain-rate regions. 
These results underscore the effectiveness of targeted control mechanisms in  reducing turbulence and enhancing flow coherence.

\section{Alignment of the lowRe case} \label{Appsec:align}
\begin{figure}%[hbt!]
\centering
\begin{minipage}[b]{1.0\textwidth}
\centering    
    \begin{subfigure}{0.45\textwidth}  
    \centering
    \includegraphics[width=0.98\textwidth]{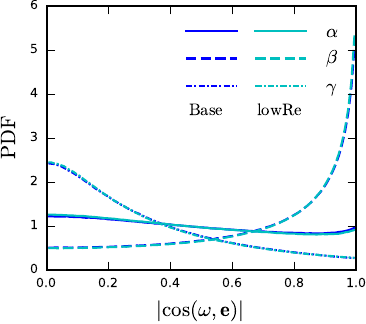}   
    \end{subfigure} 
    \begin{subfigure}{0.45\textwidth}  
    \centering
    \includegraphics[width=0.98\textwidth]{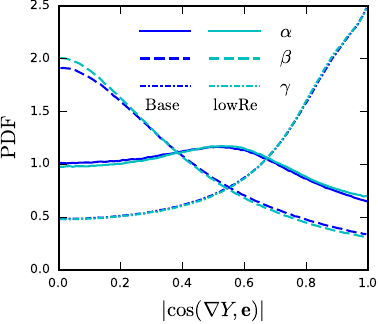}   
    \end{subfigure} 
\end{minipage}
    \caption{Comparisons of the alignment statistics between the baseline case and the lowRe cases. \label{Appfig:alignment_ori_lowRe}}
\end{figure}

The alignment of the strain-rate tensor with vorticity and the scalar gradient for both the baseline and lowRe cases is presented in Fig.~\ref{Appfig:alignment_ori_lowRe}. The results for these two cases are strikingly similar, suggesting that the significant differences in the alignment statistics observed between the baseline case and the controlled cases in Fig.~\ref{fig:alignment_stat} are not attributable to the reduction in Reynolds number. Instead, they indicate that the nonlinear interactions, as well as the energy and scalar variance transfer mechanisms, are qualitatively modified in the controlled cases.

% \section{Temporal evolution of the magnitudes of the scalar gradient budgets} \label{Appsec:dY_budgets}

% The 

\section*{Acknowledgments}
This work is supported by the National Natural Science Foundation of China (Nos.~12202270, 12102258 and 12372264). The authors also appreciate the computational support from the Centre for High Performance Computing at Shanghai Jiao Tong University.

\bibliographystyle{unsrt}
\bibliography{references}

\begin{thebibliography}{10}

\bibitem{Buzzicotti20PRL}
Michele Buzzicotti, Luca Biferale, and Federico Toschi.
\newblock Statistical properties of turbulence in the presence of a smart small-scale control.
\newblock {\em Physical Review Letters}, 124(8):084504, 2020.

\bibitem{Zhou17-1}
Ye~Zhou.
\newblock Rayleigh--taylor and richtmyer--meshkov instability induced flow, turbulence, and mixing. i.
\newblock {\em Physics Reports}, 720:1--136, 2017.

\bibitem{Zhou17-2}
Ye~Zhou.
\newblock Rayleigh--taylor and richtmyer--meshkov instability induced flow, turbulence, and mixing. ii.
\newblock {\em Physics Reports}, 723:1--160, 2017.

\bibitem{Boffetta17}
Guido Boffetta and Andrea Mazzino.
\newblock Incompressible rayleigh--taylor turbulence.
\newblock {\em Annual Review of Fluid Mechanics}, 49:119--143, 2017.

\bibitem{Livescu20}
Daniel Livescu.
\newblock Turbulence with large thermal and compositional density variations.
\newblock {\em Annual Review of Fluid Mechanics}, 52:309--341, 2020.

\bibitem{Zhouetal2021Review}
Ye~Zhou, Robin~JR Williams, Praveen Ramaprabhu, Michael Groom, Ben Thornber, Andrew Hillier, Wouter Mostert, Bertrand Rollin, S~Balachandar, Phillip~D Powell, et~al.
\newblock Rayleigh--taylor and richtmyer-meshkov instabilities: A journey through scales.
\newblock {\em Physica D: Nonlinear Phenomena}, page 132838, 2021.

\bibitem{Supernova00}
Wolfgang Hillebrandt and Jens~C Niemeyer.
\newblock Type ia supernova explosion models.
\newblock {\em Annual Review of Astronomy and Astrophysics}, 38(1):191--230, 2000.

\bibitem{Wang01}
Chih-Yueh Wang and Roger~A Chevalier.
\newblock Instabilities and clumping in type ia supernova remnants.
\newblock {\em The Astrophysical Journal}, 549(2):1119, 2001.

\bibitem{Cabot06}
William~H Cabot and Andrew~W Cook.
\newblock Reynolds number effects on rayleigh--taylor instability with possible implications for type ia supernovae.
\newblock {\em Nature Physics}, 2(8):562--568, 2006.

\bibitem{BettiHurricane16Nature}
R~Betti and OA~Hurricane.
\newblock Inertial-confinement fusion with lasers.
\newblock {\em Nature Physics}, 12(5):435, 2016.

\bibitem{zhang2018nonlinear}
H~Zhang, R~Betti, V~Gopalaswamy, R~Yan, and H~Aluie.
\newblock Nonlinear excitation of the ablative rayleigh-taylor instability for all wave numbers.
\newblock {\em Physical Review E}, 97(1):011203, 2018.

\bibitem{campbell2021direct}
EM~Campbell, TC~Sangster, VN~Goncharov, JD~Zuegel, SFB Morse, C~Sorce, GW~Collins, MS~Wei, R~Betti, SP~Regan, et~al.
\newblock Direct-drive laser fusion: status, plans and future.
\newblock {\em Philosophical Transactions of the Royal Society A}, 379(2189):20200011, 2021.

\bibitem{Ashurst96}
Wm~T Ashurst.
\newblock Flame propagation along a vortex: the baroclinic push.
\newblock {\em Combustion Science and Technology}, 112(1):175--185, 1996.

\bibitem{Keenan2014IJHE}
J.~J. Keenan, D.~V. Makarov, and V.~V. Molkov.
\newblock Rayleigh--taylor instability: Modelling and effect on coherent deflagrations.
\newblock {\em Int. J. Hydrogen Energy}, 39(35):20467--20473, 2014.

\bibitem{Sykes2021ProCI}
J.~P. Sykes, T.~P. Gallagher, and B.~A. Rankin.
\newblock Effects of rayleigh-taylor instabilities on turbulent premixed flames in a curved rectangular duct.
\newblock {\em Proc. Combust. Inst.}, 38(4):6059--6066, 2021.

\bibitem{Zhou_2024}
Ye~Zhou.
\newblock {\em Hydrodynamic Instabilities and Turbulence: Rayleigh–Taylor, Richtmyer–Meshkov, and Kelvin–Helmholtz Mixing}.
\newblock Cambridge University Press, 2024.

\bibitem{Qi2024PoF}
H.~Qi, Z.~He, A.~Xu, and Y.~Zhang.
\newblock The vortex structure and enstrophy of the mixing transition induced by rayleigh--taylor instability.
\newblock {\em Phys. Fluids}, 36(11), 2024.

\bibitem{Stanway04}
R.~Stanway.
\newblock Smart fluids: current and future developments.
\newblock {\em Materials Science and Technology}, 20(8):931--939, 2004.

\bibitem{FalconPRF17}
Eric Falcon, Jean-Claude Bacri, and Claude Laroche.
\newblock Dissipated power within a turbulent flow forced homogeneously by magnetic particles.
\newblock {\em Phys. Rev. Fluids}, 2:102601, Oct 2017.

\bibitem{Zhao22JFM}
Dongxiao Zhao, Riccardo Betti, and Hussein Aluie.
\newblock Scale interactions and anisotropy in rayleigh--taylor turbulence.
\newblock {\em Journal of Fluid Mechanics}, 930:A29, 2022.

\bibitem{ZhaoLi25}
Dongxiao Zhao, Hussein Aluie, and Gaojin Li.
\newblock Multi-scale dynamics in rayleigh-taylor turbulent mixing.
\newblock {\em Journal of Fluid Mechanics}, 802:395--436, 2025.

\bibitem{Sandoval95}
Donald~Leon Sandoval.
\newblock {\em The dynamics of variable-density turbulence}.
\newblock University of Washington, 1995.

\bibitem{YeungPope1989JFM}
Pui-Kuen Yeung and Stephen~B Pope.
\newblock Lagrangian statistics from direct numerical simulations of isotropic turbulence.
\newblock {\em Journal of Fluid Mechanics}, 207:531--586, 1989.

\bibitem{Bian20}
Xin Bian, Hussein Aluie, Dongxiao Zhao, Huasen Zhang, and Daniel Livescu.
\newblock Revisiting the late-time growth of single-mode rayleigh--taylor instability and the role of vorticity.
\newblock {\em Physica D: Nonlinear Phenomena}, 403:132250, 2020.

\bibitem{Cook2004JFM}
Andrew~W Cook, William Cabot, and Paul~L Miller.
\newblock The mixing transition in rayleigh--taylor instability.
\newblock {\em Journal of Fluid Mechanics}, 511:333--362, 2004.

\bibitem{BanerjeeAndrews2009}
Arindam Banerjee and Malcolm~J Andrews.
\newblock 3d simulations to investigate initial condition effects on the growth of rayleigh--taylor mixing.
\newblock {\em International Journal of Heat and Mass Transfer}, 52(17-18):3906--3917, 2009.

\bibitem{CabotCookNatPhys}
William~H Cabot and Andrew~W Cook.
\newblock Reynolds number effects on rayleigh--taylor instability with possible implications for type ia supernovae.
\newblock {\em Nature Physics}, 2(8):562--568, 2006.

\bibitem{MarchingCube}
William~E. Lorensen and Harvey~E. Cline.
\newblock Marching cubes: A high resolution 3d surface construction algorithm.
\newblock In {\em Proceedings of the 14th Annual Conference on Computer Graphics and Interactive Techniques}, SIGGRAPH '87, page 163–169, New York, NY, USA, 1987. Association for Computing Machinery.

\bibitem{SadekAluie18PRF}
Mahmoud Sadek and Hussein Aluie.
\newblock Extracting the spectrum of a flow by spatial filtering.
\newblock {\em Physical Review Fluids}, 3(12):124610, 2018.

\bibitem{Piomellietal91}
Ugo Piomelli, William~H Cabot, Parviz Moin, and Sangsan Lee.
\newblock Subgrid-scale backscatter in turbulent and transitional flows.
\newblock {\em Physics of Fluids A: Fluid Dynamics}, 3(7):1766--1771, 1991.

\bibitem{John2003LES}
Volker John.
\newblock {\em Large eddy simulation of turbulent incompressible flows: analytical and numerical results for a class of LES models}, volume~34.
\newblock Springer Science \& Business Media, 2003.

\bibitem{Zhao18PRF}
Dongxiao Zhao and Hussein Aluie.
\newblock Inviscid criterion for decomposing scales.
\newblock {\em Physical Review Fluids}, 3(5):054603, 2018.

\bibitem{Sequential_filter}
Dongxiao Zhao and Hussein Aluie.
\newblock Calculating spectra by sequential filtering.
\newblock {\em Journal of Renewable and Sustainable Energy}, 17(1):013303, 01 2025.

\bibitem{ZhaoAluie2023}
Dongxiao Zhao and Hussein Aluie.
\newblock Measuring scale-dependent shape anisotropy by coarse-graining: Application to inhomogeneous rayleigh-taylor turbulence.
\newblock {\em Physical Review Fluids}, 8(11):114601, 2023.

\bibitem{RistorcelliClark04}
JR~Ristorcelli and TT~Clark.
\newblock Rayleigh--taylor turbulence: self-similar analysis and direct numerical simulations.
\newblock {\em Journal of Fluid Mechanics}, 507:213--253, 2004.

\bibitem{Livescu09}
Daniel Livescu, JR~Ristorcelli, RA~Gore, SH~Dean, WH~Cabot, and AW~Cook.
\newblock High-reynolds number rayleigh--taylor turbulence.
\newblock {\em Journal of Turbulence}, (10):N13, 2009.

\bibitem{Livescu10}
D~Livescu, JR~Ristorcelli, MR~Petersen, and RA~Gore.
\newblock New phenomena in variable-density rayleigh--taylor turbulence.
\newblock {\em Physica Scripta}, 2010(T142):014015, 2010.

\bibitem{CabotZhou13}
W~Cabot and Ye~Zhou.
\newblock Statistical measurements of scaling and anisotropy of turbulent flows induced by rayleigh-taylor instability.
\newblock {\em Physics of Fluids}, 25(1):015107, 2013.

\bibitem{Schneider16JFM}
Nicolas Schneider and Serge Gauthier.
\newblock Vorticity and mixing in rayleigh--taylor boussinesq turbulence.
\newblock {\em Journal of Fluid Mechanics}, 802:395--436, 2016.

\bibitem{ashurst1987alignment}
Wm~T Ashurst, AR~Kerstein, RM~Kerr, and CH~Gibson.
\newblock Alignment of vorticity and scalar gradient with strain rate in simulated navier--stokes turbulence.
\newblock {\em The Physics of fluids}, 30(8):2343--2353, 1987.

\bibitem{livescu2021rayleigh}
Daniel Livescu, Tie Wei, and Peter~T Brady.
\newblock Rayleigh--taylor instability with gravity reversal.
\newblock {\em Physica D: nonlinear phenomena}, 417:132832, 2021.

\bibitem{NomuraPost98JFM}
Keiko~K Nomura and Gary~K Post.
\newblock The structure and dynamics of vorticity and rate of strain in incompressible homogeneous turbulence.
\newblock {\em Journal of Fluid Mechanics}, 377:65--97, 1998.

\bibitem{Meneveau2011}
Charles Meneveau.
\newblock Lagrangian dynamics and models of the velocity gradient tensor in turbulent flows.
\newblock {\em Annual Review of Fluid Mechanics}, 43:219--245, 2011.

\bibitem{JohnsonWilczek2024}
Perry~L Johnson and Michael Wilczek.
\newblock Multiscale velocity gradients in turbulence.
\newblock {\em Annual Review of Fluid Mechanics}, 56(1):463--490, 2024.

\bibitem{White08ARFM}
Christopher~M White and M~Godfrey Mungal.
\newblock Mechanics and prediction of turbulent drag reduction with polymer additives.
\newblock {\em Annu. Rev. Fluid Mech.}, 40(1):235--256, 2008.

\bibitem{Procaccia2008RMP}
Itamar Procaccia, Victor~S L’vov, and Roberto Benzi.
\newblock Colloquium: Theory of drag reduction by polymers in wall-bounded turbulence.
\newblock {\em Reviews of Modern Physics}, 80(1):225--247, 2008.

\bibitem{Xi2019PoF}
Li~Xi.
\newblock Turbulent drag reduction by polymer additives: Fundamentals and recent advances.
\newblock {\em Physics of Fluids}, 31(12), 2019.

\bibitem{XuXi23JFM}
Ao~Xu, Ben-Rui Xu, and Heng-Dong Xi.
\newblock Wall-sheared thermal convection: heat transfer enhancement and turbulence relaminarization.
\newblock {\em Journal of Fluid Mechanics}, 960:A2, 2023.

\end{thebibliography}
\end{document}